\documentclass[sigconf, screen]{acmart}

\usepackage{tcolorbox}
\tcbuselibrary{skins, breakable}  

\usepackage{listings}
\usepackage{xcolor}
\usepackage[table,dvipsnames]{xcolor}
\usepackage{multirow}

\usepackage[utf8]{inputenc}
\usepackage{amsmath}

\newenvironment{unbreakablelisting}
  {\par\noindent\begin{minipage}{\columnwidth}}
  {\end{minipage}\par}

\usepackage[ruled,vlined]{algorithm2e}
\SetKw{Continue}{continue}
\usepackage{graphicx}
\usepackage{subcaption}
\usepackage{svg}
\usepackage{transparent}
\usepackage{enumitem}

\usepackage{url}

\usepackage{array}

\usepackage{booktabs}
\usepackage{pifont}
\usepackage{threeparttable}
\usepackage{makecell}

\usepackage{hyperref}

\definecolor{diffaddbg}{RGB}{230, 255, 236} 
\definecolor{diffrembg}{RGB}{255, 235, 232} 
\definecolor{keywordblue}{RGB}{0, 102, 204}
\definecolor{typeblue}{RGB}{0, 128, 128}

\newcommand{\tool}{FeatureFuzz\xspace}
\newcommand{\toolft}{$FF_{Group}$\xspace}
\newcommand{\toolrandomcov}{$FF_{RandCov}$\xspace}
\newcommand{\toolrandom}{$FF_{Rand}$\xspace}
\newcommand{\toolsingle}{$FF_{Single}$\xspace}
\newcommand{\bugft}{SFT\xspace}

\newcommand{\topic}[1]{\textbf{#1}}

\newcommand{\metamut}{MetaMut\xspace}
\newcommand{\mutfall}{Mut4All\xspace}
\newcommand{\issuemut}{Issuemut\xspace}
\newcommand{\yarpgen}{YARPGen\xspace}
\newcommand{\csmith}{Csmith\xspace}
\newcommand{\fuzzfall}{Fuzz4All\xspace}
\newcommand{\creal}{Creal\xspace}
\newcommand{\legofuzz}{LegoFuzz\xspace}

\newcommand{\extractllm}{ExtractLLM\xspace}
\newcommand{\instantiatellm}{InstanLLM\xspace}
\newcommand{\groupllm}{GroupLLM\xspace}

\newcommand{\numberReportedBug}{122\xspace}
\newcommand{\numberReportedBugLLVM}{62\xspace}
\newcommand{\numberReportedBugGCC}{60\xspace}

\newcommand{\numberConfirmBug}{106\xspace}
\newcommand{\numberConfirmBugLLVM}{60\xspace}
\newcommand{\numberConfirmBugGCC}{46\xspace}

\newcommand{\numberAssignedBug}{31\xspace}
\newcommand{\numberAssignedBugLLVM}{16\xspace}
\newcommand{\numberAssignedBugGCC}{15\xspace}

\newcommand{\numberDuplicateBug}{4\xspace}
\newcommand{\numberDuplicateBugLLVM}{2\xspace}
\newcommand{\numberDuplicateBugGCC}{2\xspace}

\newcommand{\numberFrontEndBug}{79\xspace}
\newcommand{\numberFrontEndBugLLVM}{32\xspace}
\newcommand{\numberFrontEndBugGCC}{47\xspace}

\newcommand{\middleBackendBug}{43\xspace}

\newcommand{\numberMiddleEndBug}{26\xspace}
\newcommand{\numberMiddleEndBugLLVM}{17\xspace}
\newcommand{\numberMiddleEndBugGCC}{9\xspace}

\newcommand{\numberBackEndBug}{17\xspace}
\newcommand{\numberBackEndBugLLVM}{13\xspace}
\newcommand{\numberBackEndBugGCC}{4\xspace}

\newcommand{\numberValidBug}{48\xspace}

\newcommand{\numberReleaseBuildBug}{64\xspace}
\newcommand{\numberNonReleaseBuildBug}{58\xspace}
\newcommand{\numberReleaseAssertionBug}{22\xspace}
\newcommand{\numberReleaseFPEBug}{1\xspace}
\newcommand{\numberReleaseMemorySafetyBug}{9\xspace}
\newcommand{\numberReleaseRuntimeFaultBug}{32\xspace}

\copyrightyear{2026}
\acmYear{2026}
\setcopyright{cc}
\setcctype{by}
\acmConference[CCS '26]{Proceedings of the 2026 ACM SIGSAC Conference on Computer and Communications Security}{November 15--19, 2026}{The Hague, Netherlands}
\acmBooktitle{Proceedings of the 2026 ACM SIGSAC Conference on Computer and Communications Security (CCS '26), November 15--19, 2026, The Hague, Netherlands}
\acmDOI{10.1145/3830454.3832578}
\acmISBN{979-8-4007-2871-6/2026/11}
\author{Xingbang He}
\affiliation{
\institution{State Key Laboratory of Novel Software Technology, Nanjing University}
\city{Nanjing}
\country{China}
}

\author{Yuanwei Chen}
\affiliation{
\institution{State Key Laboratory of Novel Software Technology, Nanjing University}
\city{Nanjing}
\country{China}
}

\author{Hao Wu}
\affiliation{
\institution{State Key Laboratory of Novel Software Technology, Nanjing University}
\city{Nanjing}
\country{China}
}

\author{Jikang Zhang}
\affiliation{
\institution{Institute of Dataspace, Hefei Comprehensive National Science Center}
\city{Hefei}
\country{China}
}

\author{Zicheng Wang}
\affiliation{
\institution{Honor Device Co., Ltd}
\city{Beijing}
\country{China}
}

\author{Ligeng Chen}
\affiliation{
\institution{Honor Device Co., Ltd}
\city{Beijing}
\country{China}
}

\author{Junjie Peng}
\affiliation{
\institution{State Key Laboratory of Novel Software Technology, Nanjing University}
\city{Nanjing}
\country{China}
}

\author{Haiyang Wei}
\affiliation{
\institution{State Key Laboratory of Novel Software Technology, Nanjing University}
\city{Nanjing}
\country{China}
}

\author{Yi Qian}
\affiliation{
\institution{State Key Laboratory of Novel Software Technology, Nanjing University}
\city{Nanjing}
\country{China}
}

\author{Tiantai Zhang}
\affiliation{
\institution{State Key Laboratory of Novel Software Technology, Nanjing University}
\city{Nanjing}
\country{China}
}

\author{Linzhang Wang}
\affiliation{
\institution{State Key Laboratory of Novel Software Technology, Nanjing University}
\city{Nanjing}
\country{China}
}

\author{Bing Mao}
\authornote{*Corresponding author. maobing@nju.edu.cn.}
\affiliation{
\institution{State Key Laboratory of Novel Software Technology, Nanjing University}
\city{Nanjing}
\country{China}
}

\begin{document}

\title{FeatureFuzz: Treating Bug-prone Features as Core Primitives for Compiler Fuzzing}

%

\renewcommand{\shortauthors}{Xingbang He et al.}

\begin{abstract}
CI/CD pipelines often execute build and test tasks with high privileges, making the compilation stage security-sensitive. 
In this setting, malicious pull requests may exploit carefully crafted source code to trigger arbitrary code execution or excessive resource consumption during compilation, making compiler vulnerabilities security-relevant. 
Recent work has leveraged bug-prone knowledge from historical bugs to drive compiler exploration and bug discovery. 
However, many compiler bugs depend on program semantics rather than syntax, and existing approaches either rely on syntactic mutation or encode this knowledge in model parameters, making it difficult to preserve and recombine the semantics that actually trigger compiler bugs.
We propose \tool, a compiler fuzzer that explicitly reuses bug-prone semantics through features and feature groups. 
\tool extracts these bug-prone semantics from historical bugs as features, synthesizes them into coherent feature groups, and instantiates them into programs for compiler fuzzing. 
On GCC and LLVM, over 24-hour campaigns, \tool uncovers 167 unique crashes, 2.78$\times$ as many as the second-best fuzzer. 
Over a 72-hour fuzzing campaign, \tool identifies \numberReportedBug bugs, of which \numberConfirmBug have been confirmed by compiler developers as zero-days. 
Among reported bugs, \numberReleaseBuildBug affect release-build compilers, including \numberReleaseMemorySafetyBug memory-corruption bugs, highlighting the security relevance of compiler fuzzing in CI/CD settings.

\end{abstract}

%
%

\ccsdesc[300]{Security and privacy~Software and application security}
%
\keywords{Compiler, Compiler Fuzzing, Compiler Bugs}


\maketitle

\section{Introduction}

Continuous Integration and Continuous Deployment (CI/CD) pipelines have become a basic part of modern software engineering~\cite{ARGUS2023}. 
They routinely automate building, testing and releasing 
through CI/CD frameworks such as Travis CI~\cite{travisci} and GitLab CI~\cite{gitlabci}.
They often execute on runners with high privileges because they may access credentials, sensitive logs, and release permissions~\citeN{sharma2025compliancecicdrunners, nisimi2024leakycli, praetoriantensorflowrunner}. 

CI/CD pipelines are security-sensitive, and compromising them can lead to credential theft, data theft, and malware propagation~\citeN{owaspcicd, ehrman2025cicdsecurity}. 
The TensorFlow~\cite{tensorflow} self-hosted runner attack~\cite{praetoriantensorflowrunner} illustrates this risk, in which a malicious pull request gained remote code execution (RCE) on the runner and enabled malicious releases to the official TensorFlow repository. 
As a core stage of CI/CD pipelines, compilers form a security-relevant attack surface. 
Maliciously crafted programs can exploit compiler vulnerabilities to trigger arbitrary code execution or unconstrained resource consumption on CI/CD runners~\cite{gccsecurity}. 
These risks make rigorous compiler testing necessary. 

Fuzzing has emerged as one of the most effective techniques for systematically uncovering compiler defects~\cite{CompilerTestingSurvey2020}, 
and has exposed thousands of vulnerabilities in mainstream compilers such as GCC and LLVM~\cite{CompilerFuzzingHowMuchDoesItMatter}.
Modern compiler fuzzing primarily follows three trajectories: generation-based, mutation-based, and approaches based on large language models (LLMs). 
Generation-based fuzzers~\citeN{CsmithPLDI11,YARPGen2020,YARPGenv22023} synthesize programs from hand-crafted grammars, prioritizing syntactic validity but struggling to cover bug-prone semantic states. 
Mutation-based fuzzers~\citeN{EMIPLDI14, MetaMut2024, Mut4All2025, GrayC2023} iteratively transform seed programs, yet often inadvertently disrupt delicate semantic invariants required to trigger deep bugs. 
More recently, LLM-based fuzzers~\citeN{fuzz4all2024, FuzzGPT2024, MirrorFuzzr2025, ClozeMaster2025} leverage the vast training corpus of LLMs for program synthesis. 
While expressive, these LLM-based fuzzers encode bug-prone knowledge in model parameters, making it difficult to explicitly preserve and recombine the semantics needed to trigger compiler defects.

To enhance program diversity, recent research has started using bug-prone patterns from historical bug reports~\citeN{Mut4All2025, Issue2mut2025, ASMFUZZ2024, ClozeMaster2025}. These methods typically transform bug-prone logic into reusable artifacts, such as specialized mutators~\citeN{Mut4All2025, Issue2mut2025,ASMFUZZ2024, Grammarinator2019, LangFuzz2012} or fine-tuned model weights~\citeN{ClozeMaster2025, DSmith2020, DeepSmith2018}. 
However, they suffer from a fundamental limitation that we call \textit{Semantic Collapse}: bug-prone semantics are no longer explicit, reusable, and recombinable primitives for fuzzing, because they are either reduced to shallow syntactic transformations or buried in the model's latent parameters.
The semantics that truly trigger compiler defects, such as aliasing relationships, type constraints, and subtle data dependencies, are either collapsed into shallow syntactic transformations or remain inaccessible within a model's black box. 
Consequently, existing fuzzers fail to faithfully preserve and systematically recombine the underlying logical conditions that drive compilers into inconsistent states, ultimately constraining the fuzzer's capacity to fully explore the semantic space. \looseness=-1

The core challenge is the absence of an explicit representation that can faithfully preserve bug-prone semantics while enabling flexible recombination.
Semantics describe the logical relationships between program elements independently of their syntactic implementation. 
Without a representation aligned with these underlying semantics, existing efforts to reuse bug-prone logic inevitably devolve into syntax-bound transformations or remain trapped in the model's parameters. 
These implementations fail to capture the semantics inherent in bug-prone logic, which fundamentally limits the diversity of generated programs.

Bugs are generally rooted in semantics rather than specific syntactic structures. 
Numerous reports in the compiler community reveal duplicated bug instances: although the bug-triggering programs are syntactically different, they share common semantics that lead to the same bug.
This suggests that we can represent such semantics using natural language, rather than focusing on specific code formats. 
Furthermore, natural language serves as an expressive and practical proxy for representing these semantics.
Natural language is suited to describe high-level program relationships, such as the complex invariants between base and derived classes, without the prohibitive overhead of formal verification or manual modeling.
Traditionally, capturing these semantics has required intensive domain expertise and manual engineering.
For example, ensuring the relationship between an array index and its length often necessitates either hand-crafted constraints, as seen in YARPGen~\cite{YARPGen2020}, or specialized runtime instrumentation, as used in UBFuzz~\cite{UBFuzz2024}.
Given the sheer variety of semantics in real-world bug histories, manually formalizing each pattern is unscalable. Fortunately, the emergence of LLMs bridges this gap, as they can directly translate high-level natural language descriptions into executable code~\cite{chen2021evaluatinglargelanguagemodels}. By adopting natural language as an intermediate semantic representation, we can explicitly preserve and systematically recombine bug-prone logic, thereby enabling more effective compiler fuzzing. \looseness=-1

Guided by this insight, we propose \tool, a compiler fuzzer that shifts the generation paradigm from syntax-level mutation to feature-group construction. In our context, a \textbf{Feature} is defined as a dual-component primitive: (1) a natural language description of a bug-prone invariant that is decoupled from concrete program syntax, such as an out-of-bounds array access; and (2) a code witness providing a concrete implementation of that invariant. \tool treats \textit{feature groups} as the central semantic objects and operates through a three-stage pipeline:
\textit{(1) Extraction:} Using \extractllm, it distills reusable features from historical bug reports and organizes them into a feature pool, transforming program-specific bug-triggering logic into abstract semantic primitives.
\textit{(2) Synthesis:} It employs a fine-tuned LLM (\groupllm) to combine multiple features into a coherent feature group, introducing auxiliary ``glue'' semantics to bridge the implicit dependencies between isolated logic fragments.
\textit{(3) Instantiation:} It leverages \instantiatellm to instantiate each synthesized feature group as a syntactically valid program that jointly satisfies the group's semantic constraints.

We evaluate \tool on GCC and LLVM.  
Over 24-hour fuzzing campaigns, \tool outperforms state-of-the-art fuzzers \metamut~\cite{MetaMut2024}, \mutfall~\cite{Mut4All2025}, \fuzzfall~\cite{fuzz4all2024}, \yarpgen~\cite{YARPGen2020} and \legofuzz~\cite{LegoFuzz2025} in terms of both coverage and unique crashes. 
\tool uncovers 167 crashes, which is 2.78$\times$ as many as the second-best fuzzer, \metamut.
Through a 72-hour fuzzing campaign by \tool, we reported \numberReportedBug bugs to GCC and LLVM developers. \numberConfirmBug of these have already been confirmed by compiler developers, and \middleBackendBug are found in the middle-end or back-end compiler components. Among reported bugs, \numberReleaseBuildBug affect release-build compilers, including \numberReleaseMemorySafetyBug memory-corruption bugs, indicating security relevance.
Additionally, \numberValidBug bugs are triggered by compilable programs that could occur in real-world programming.
Our results show that explicitly exploring feature combinations is an effective and practical approach to stress-testing modern compilers.
In summary, this paper makes the following contributions:
\begin{itemize}[leftmargin=*, topsep=3pt, itemsep=2pt]
    \item We introduce bug-prone features, decoupled representations that combine natural language descriptions with code witnesses. With features that expressively and precisely capture semantics in bug-prone logic and are reusable, we enhance the diversity of test programs for compiler testing.

    \item We design and implement \tool, a feature-based fuzzing framework. It extracts features from historical bugs, composes them into coherent feature groups using a fine-tuned LLM, and instantiates these groups into executable test programs.
    
    \item Our evaluation on GCC and LLVM shows that \tool significantly outperforms existing baselines in both coverage and bug discovery, uncovering 167 unique crashes. \tool reported \numberReportedBug new bugs, including \numberConfirmBug developer-confirmed zero-days. \looseness=-1
\end{itemize}

\section{Threat Model}
\label{sec:threat_model}

We consider CI/CD compilation pipelines executed on security-sensitive runners, where the execution environment may hold sensitive information or high-risk privileges, such as repository credentials, signing keys, or release permissions.

\noindent \textbf{Attacker capabilities:}
We assume that an attacker can control source code in the pipeline and trigger its execution, for example by submitting a pull request that causes compilation to run on a security-sensitive runner.

\noindent \textbf{Impact:}
A carefully crafted source program may threaten the availability of the CI/CD system by driving the compiler to consume excessive resources, thereby delaying or terminating other workflows on the same runner. More seriously, compiler bugs may have security implications for the compilation stage of CI/CD pipelines, and maliciously and carefully crafted programs may achieve remote code execution~\cite{gccsecurity}.

Under this threat model, more effective compiler testing is needed. However, existing fuzzers still struggle to faithfully preserve and recombine bug-prone semantics. 

\section{Motivating Example}
\label{sec:motivation}

\begin{unbreakablelisting}
\begin{lstlisting}[language=C, 
caption={A valid program triggers an internal compiler error (ICE) in GCC-15 but passes in other GCC versions.}, 
label=lst:motivation, 
basicstyle=\footnotesize\ttfamily,
]
void func() {}
int main() {
    void *p = (void*)&func;
    double w_1[2] = {1, 2};
    int i = 2;
    if (w_1[i] < 1.0) { 
        goto *p;
    }
    return 0;
}
\end{lstlisting}
\end{unbreakablelisting}

Listing~\ref{lst:motivation} illustrates a critical bug we discovered in GCC-15.2 (at the \texttt{-O3} optimization level). This bug triggers an internal compiler error (ICE), yet remains undetected by state-of-the-art C/C++ fuzzers. Through root-cause analysis, we identified that triggering this defect requires the simultaneous satisfaction of four features:

\begin{enumerate}[leftmargin=*]
    \item[\textbf{F1}] An array access is used in a conditional expression (e.g., the \texttt{if} statement in line~6).
    \item[\textbf{F2}] An index variable whose value exceeds the array bounds (e.g., the variable \texttt{i} in line~6).
    \item[\textbf{F3}] A computed non-local goto whose target is a function address (e.g., the \texttt{goto *p} statement in line~7).
    \item[\textbf{F4}] A goto statement located in either branch of an if--else statement (e.g., the \texttt{goto *p} is in the \texttt{if} branch).
\end{enumerate}

\noindent \textbf{The Challenges of Bug Triggering.} Constructing a program to trigger such a bug is non-trivial due to two fundamental challenges:

\begin{itemize}[leftmargin=2em]
    \item[\textbf{C1}] \textbf{Semantic Constraints.} Many bug-triggering conditions, such as \textbf{F2}, are inherently flow-sensitive. They depend on the runtime values of variables rather than static syntax. While a grammar-based fuzzer can easily generate an array access \texttt{w\_1[i]}, ensuring that \texttt{i} consistently exceeds the bounds of \texttt{w\_1} requires global constraints that are difficult to formalize in a context-free grammar or a mutation engine.
    \item[\textbf{C2}] \textbf{Implicit Dependencies.} The features in Listing~\ref{lst:motivation} are not independent; they are logically interleaved. For instance, there is an implicit dependency between \textbf{F1} and \textbf{F4}: the array-based condition in \textbf{F1} must act as the predicate for the branch containing the \texttt{goto} in \textbf{F4}. Triggering the bug requires more than the presence of these features; it requires their coordinated orchestration.
\end{itemize}

\noindent \textbf{Limitations of State-of-the-art.} 
Existing compiler fuzzing trajectories fail to address these challenges.
\textit{First}, mutation-based fuzzers~\cite{MetaMut2024, Mut4All2025, GrayC2023} typically perform localized structural edits. When a seed program only possesses a subset of the required features, attempting to inject the remaining ones often disrupts existing semantic invariants~\cite{FuzzerMutatorPerformance2024}. This is because these tools cannot coordinate mutations across multiple program locations to satisfy the required feature combination~\cite{Issue2mut2025}. 
\textit{Second}, LLM-based generators~\cite{MirrorFuzzr2025, ClozeMaster2025, FuzzGPT2024, fuzz4all2024} struggle to synthesize such multi-feature combinations because their training corpora typically present these features in a decoupled state. Without explicit guidance, the model struggles with the long-tail joint distribution of these features, where the individual features may be common, but their joint occurrence is rare, making the bug difficult to trigger. \looseness=-1

\begin{figure*}[ht]
  \centering
  \includegraphics[width=1\textwidth]{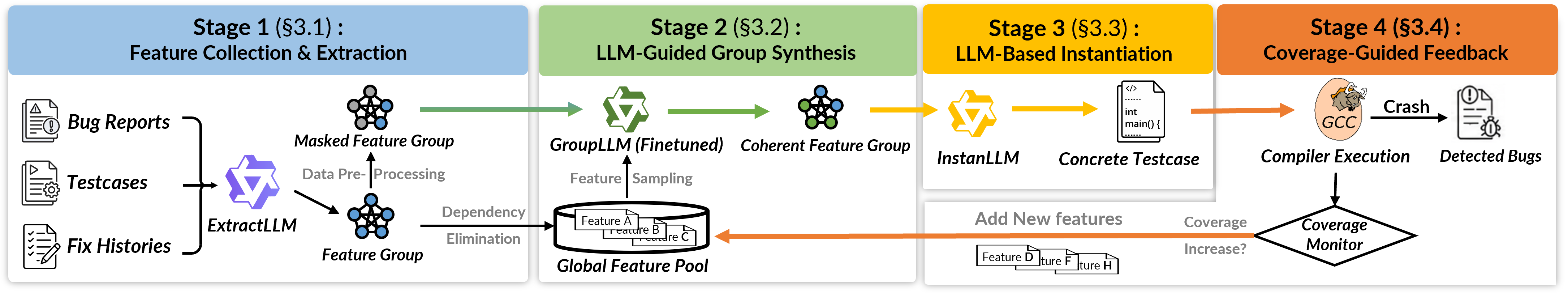}
  \caption{Overview of \tool.}
  \label{fig:overview_png}
\end{figure*}

\noindent \textbf{Our Observations and Insights.} 
The inability of existing tools reveals a fundamental gap: effective bug triggering requires moving beyond localized syntactic edits toward an explicit understanding of the causal logic that binds isolated program elements. 
To bridge this gap, we offer two key insights:

\begin{itemize}[leftmargin=2em, topsep=3pt, partopsep=3pt]
\item[I1] \textbf{Feature Decoupling.} We observe that the bug-triggering potential of features like \textbf{F3} is independent of their concrete syntax; the \texttt{goto} could reside in either branch or use different variable names. By abstracting features into a decoupled representation, we can escape the syntax-bound limitation of mutators. This abstraction allows us to treat complex constraints (\textbf{C1}) as high-level properties rather than hard-coded templates.

\item[I2] \textbf{Recombination Bridge.} While abstraction breaks explicit syntactic links, it reveals that the dependencies between features (\textbf{C2}) are logical roles. We suggest that natural language is a practical proxy to describe these roles (e.g., ``this feature provides a predicate''). By using an LLM to reason over these descriptions, we can glue isolated features back together, resolving the implicit dependencies that previously caused semantic collapse.
\end{itemize}

Guided by these insights, the next section presents the design of \tool, a framework dedicated to the systematic extraction and recombination of features.


\section{\tool Design}
\label{sec:approach}



We introduce \tool, a compiler fuzzer that shifts the paradigm from syntax-level mutation to feature composition. Driven by the insight that complex bugs are often triggered by interdependent program properties, \tool aggregates discrete features into unified \textit{feature groups}, which are then realized as programs. \looseness=-1

Figure~\ref{fig:overview_png} illustrates the four-stage architecture of \tool:
\begin{enumerate}[leftmargin=*, topsep=5pt, partopsep=3pt]
\item \textbf{Feature Extraction (\S\ref{sec:feature-extraction}):} \tool utilizes \extractllm to distill feature groups from historical compiler bug reports and their corresponding proof-of-concept (PoC) programs. These groups are then decomposed into individual, reusable features to populate a global \textit{feature pool}.
\item \textbf{Feature Group Synthesis (\S\ref{sec:feature-group-synthesis}):} To explore the combination of semantics, \tool samples features from the pool and employs a fine-tuned LLM (\groupllm) to synthesize them into a coherent \textit{feature group}. This stage introduces glue semantics to ensure logical consistency and inter-feature dependency.
\item \textbf{Test Case Instantiation (\S\ref{sec:testcase-generation}):} \tool leverages \instantiatellm to transform abstract feature groups into syntactically valid and semantically consistent programs, ensuring that multiple features are combined within a unified execution context.
\item \textbf{Coverage-Guided Feedback (\S\ref{sec:coverage-guided-feedback}):} \tool maintains an evolutionary loop where feature groups that trigger novel code coverage are prioritized, and their constituent features are rewarded or refined in the pool for future iterations. During compiler execution, \tool detects bug candidates by monitoring compiler crashes, which are then deduplicated and analyzed.
\end{enumerate}

\subsection{Feature Extraction}
\label{sec:feature-extraction}

\begin{figure}[H] 
\begin{tcolorbox}[
  colback=cyan!8!white,      
  colframe=black!65,   
  boxrule=1.5pt,       
  arc=1mm,             
  left=0em, right=0em, top=0em, bottom=0em,  
  enhanced, 
  title=Feature Extraction,
  fonttitle=\bfseries, 
]

\textbf{[Input]}
A bug-triggering program, its corresponding bug report, and the root cause of the bug described in the fix history.

\textbf{[Task]}
Extract and describe key features that a program need in order to reproduce the bug described in the input.

\textbf{[Steps]}
1. Read the bug report and the root cause description to understand the observed failure and why the bug occurs.\\
2. Analyze the bug report, root cause, and the program together to determine the key features relevant to triggering the issue.\\
3. Extract the key features from the input.

\textbf{[OutputExample]}
The code should include a value computed inside a loop that remains unchanged throughout the iterations and is subsequently used after the loop, e.g., \texttt{for (i = 0; i < n; i++) \{ x = global\_var; \} result = x + i;}.
\end{tcolorbox}
\vspace{-10pt}
\caption{The feature extraction prompt} 
\label{fig:feature-extraction-box}     
\end{figure}

In our context, a \textbf{Feature} is defined as a dual-component primitive: (1) a natural language description of a bug-prone invariant, and (2) a code witness providing a concrete implementation of that invariant. The code witness also preserves syntax-level triggering details that may be lost in pure natural-language abstraction, such as exact data types, qualifiers, or attributes. The following example presents a feature relevant to loop optimization:

\textit{Description}: The code should include a value computed inside a loop that remains unchanged throughout the iterations and is subsequently used after the loop. \textit{Code}: \texttt{for (i = 0; i < n; i++) \{ x = global\_var; \} result = x + i;}
 
The code of this feature helps \instantiatellm understand and instantiate it within programs.
Such a feature may trigger a compiler optimization that hoists \texttt{x} outside of the loop.

We deliberately decouple features from specific syntax for three reasons:
\textbf{(1) Expressiveness:} Natural language can describe complex, flow-sensitive properties (e.g., aliasing or value ranges) that are difficult to encode via fixed Abstract Syntax Tree (AST) templates.
\textbf{(2) Precision:} By abstracting away implementation details (e.g., specific variable names or formatting), \tool focuses the search space on the core logic known to stress compiler backends.
\textbf{(3) Reusability:} Abstracted features are modular, allowing them to be seamlessly recombined with features extracted from entirely different program contexts.

As shown in Figure~\ref{fig:feature-extraction-box}, \extractllm acts as a bridge, translating these into structured feature groups whose features capture the key semantics that trigger bugs while remaining independently represented, without explicit cross-references.
By anchoring these features to real-world defects, we ensure that our fuzzer is biased toward bug-prone regions of the compiler's optimization space, a strategy proven effective in prior studies~\cite{OrionDLFuzz2025, ComFuzz2023}.
These extracted groups serve a dual purpose: they populate the feature pool and act as training data to fine-tune \groupllm, teaching it how to bridge disparate features into logically sound programs.

\subsection{Feature Group Synthesis}
\label{sec:feature-group-synthesis}

To construct a high-quality feature pool, \tool performs semantic extraction 
by abstracting bug-prone invariants from
on historical artifacts, including bug-triggering programs, bug reports, and commit fix histories.
These artifacts can be collected from compiler-specific bug report websites and open-source repositories, such as the GCC Bugzilla~\footnote{\url{https://gcc.gnu.org/bugzilla/}} and the official GCC repository~\footnote{\url{https://gcc.gnu.org/git.html}}, and provide the ground truth for why a specific logic fragment causes a compiler to fail.
We build the \extractllm with a pre-trained LLM and the following prompt (Figure~\ref{fig:feature-extraction-box}). The prompt instructs the model to identify not just the syntax of a PoC, but the underlying causal conditions.

With a populated feature pool, the challenge shifts to constructing coherent feature groups where constituent features exhibit logical interdependence.
Naive random sampling from the global pool often results in semantic fragmentation, where features extracted from disparate bug reports lack the implicit dependencies (e.g., shared data flow or control structures) necessary for a unified program.
Forcing such unrelated features into a single program often leads to either syntactically invalid code or disjointed logic that fails to exercise complex compiler optimizations.

Furthermore, relying solely on sampling from a fixed pool restricts the fuzzer to a static search horizon. While the number of potential combinations is vast, the semantic boundaries remain confined to previously observed bug patterns, limiting the fuzzer's ability to explore novel or zero-day compiler behaviors.

To resolve the tension between semantic cohesion and search diversity, we employ a fine-tuned LLM, \groupllm, to synthesize auxiliary features that glue the sampled features into a coherent group. 
These synthesized features serve two primary functions:
\textit{First}, instead of modifying the original bug-prone features, which could inadvertently sanitize the very logic that triggers a defect, \groupllm introduces glue semantics. These auxiliary features provide the necessary scaffolding (e.g., common variable declarations or interface wrappers) to bridge the gap between isolated features.
\textit{Second}, by leveraging the LLM's vast pre-trained knowledge of programming semantics, \groupllm can propose novel features that are logically related to the sampled set but not present in the historical pool, effectively expanding the fuzzer's reach into unexplored compiler states.

We build the \groupllm by fine-tuning a pre-trained LLM for this task because identifying implicit dependencies in natural language requires a deep, contextual understanding of program logic.
Since our features deliberately omit low-level implementation details to maintain abstraction, the model must reason about how high-level concepts (e.g., a ``non-local goto'' and an ``array-bound constraint'') can be logically coupled.

\noindent \textbf{Fine-tuning Method.} \groupllm is fine-tuned using the original feature groups distilled from historical bugs (as described in \S\ref{sec:feature-extraction}). 
We employ a \textbf{masked semantic prediction} objective during training.
Within the collected feature groups, these features no longer have explicit relationships based on concrete code, but rather exhibit implicit relationships. 
Therefore, they possess no inherent order. 
This order independence allows us to randomly mask parts of the features in a group to generate a large amount of training data.
For each feature group, we randomly partition its features into non-empty input and target subsets, while controlling the number of generated pairs according to group size to augment sparse groups without allowing large groups to dominate the training corpus.
Detailed data construction and fine-tuning parameters are provided in Appendix~\ref{appendix:groupllm}.

Because these training groups are derived from real-world, functional programs, they embody high semantic cohesion. 
This semantic cohesion ensures that the masked prediction task is not a stochastic prediction, but rather a process of dependency supplementation.
By learning to predict masked features, \groupllm internalizes the underlying causal relationships and structural patterns common in bug-triggering programs.
Consequently, during the synthesis phase, the model treats the randomly sampled features as a prompt and generates the most probable missing features to maximize the group's overall cohesion and diversity.

\subsection{Program Instantiation}
\label{sec:testcase-generation}

\begin{figure}[H] 
\centering 
\begin{tcolorbox}[
  colback=cyan!8!white,      
  colframe=black!65,   
  boxrule=1.5pt,       
  arc=1mm,             
  left=0em, right=0em, top=0em, bottom=0em,  
  enhanced,
  title=Program Instantiation,
  fonttitle=\bfseries
]

\textbf{[Task]}
Given a set of features, generate a single C/C++ program that satisfies all features and stresses compiler edge cases.

\textbf{[Input]}
A set of features where each feature specifies a semantic that must be satisfied by the generated program.

\textbf{[Instructions]}
1. All features must be satisfied within a single program.
2. Dependencies among features are \emph{implicit} and should be realized through appropriate control-flow and/or data-flow structures in the generated code.
3. The generated code must be valid and compilable under the language specification.
\end{tcolorbox}
\vspace{-10pt}
\caption{The prompt for program instantiation} 
\label{fig: Program-Instantiation-template}
\end{figure}

The next stage of the pipeline is program instantiation, where \instantiatellm transforms the synthesized feature groups into executable programs via a task-specific prompt (Figure~\ref{fig: Program-Instantiation-template}).
This process entails two tasks: 
(1) ensuring the joint satisfaction of all individual features within the group, and 
(2) materializing the implicit dependencies synthesized by \groupllm into concrete program structures.

Unlike template-based generators that fill predefined slots, \instantiatellm performs a holistic synthesis to satisfy multiple semantic constraints simultaneously.
A program is considered a successful instantiation only if all features in the group are jointly realized. To facilitate this, \instantiatellm leverages the code witnesses (instances) accompanying each feature in the pool. These witnesses provide the model with concrete implementation patterns that it can adapt and integrate into a unified context.

A critical challenge in instantiation is the explicit realization of latent dependencies.
Because features are represented as abstract natural language rather than fixed AST fragments, their interrelations, such as data-flow across functions or control-flow nesting, are implicit. Traditional rule-based or template-driven strategies struggle to resolve these dependencies reliably. Consequently, \instantiatellm is tasked with dynamically constructing the following structures:
\textit{(1) Control-Flow Integration:} The model orchestrates the execution order, branching logic, and nesting depth required to link disparate features (e.g., placing a memory access feature inside a specific loop invariant). \textit{(2) Data-Flow Synthesis:} The model establishes value propagation, variable sharing, and state dependencies across statements, ensuring that the output of one feature correctly serves as the predicate or operand for another.

These structures are not hard-coded; instead, they emerge from the LLM's reasoning over the requested feature group. By guiding the generation with explicit semantic constraints, \tool avoids the structural redundancy and hallucinated logic often seen in unconstrained LLM-based fuzzers~\citeN{FuzzGPT2024,fuzz4all2024,ClozeMaster2025}. This constraint-driven approach ensures that the resulting programs are not only syntactically valid but also semantically dense, maximizing the likelihood of stressing complex compiler optimizations.

\subsection{Coverage-guided Feedback Fuzzing}
\label{sec:coverage-guided-feedback}

To iteratively refine the quality of the feature pool, \tool incorporates a coverage-guided feedback loop. This mechanism treats code coverage as a proxy for the effectiveness of the features synthesized by \groupllm.
By rewarding features that contribute to novel coverage, \tool transitions from a static corpus to an evolving feature set. 

\tool utilizes coverage feedback to quantify the effectiveness of feature groups and resolve the credit assignment problem.
When an instantiated program triggers novel code coverage, it indicates a successful exploration of previously unreached compiler states. While a single program is merely one concrete instance of a feature group, we consider the coverage gain as an indicator of the \textit{latent potential} of the entire group.
Therefore, \tool optimistically attributes the exploratory success to the feature group as a whole.

Specifically, the glue features introduced by \groupllm, which represent newly synthesized semantic logic, are only graduated to the global feature pool if their parent group yields a coverage increase. Isolating the precise contribution of a single feature within a complex program is a non-trivial challenge in program analysis. To address this, we adopt an optimistic promotion strategy: any observed coverage gain validates the efficacy of the newly introduced features, which are then prioritized in subsequent fuzzing iterations to further explore the surrounding semantic space.

\begin{algorithm}[h]
  \caption{Coverage-guided Feedback Fuzzing Loop}
  \label{alg:feature-selection}

  \SetKwInOut{Input}{Input}
  \SetKwInOut{Output}{Output}
  \SetKwComment{Comment}{// }{}
  
  \Input{
     Global feature pool $\mathcal{F}$, group size $k$, \\
     coverage sampler $\mathsf{Cov}$, \groupllm, \instantiatellm
  }
  
  \BlankLine
  Novel features queue $\mathcal{N} \leftarrow \emptyset$\;
  $C \leftarrow \mathsf{Cov}.\text{sample}()$\;
  
  \Repeat{timeout}{
      $k_N \leftarrow \text{UniformInt}(0, \min(k, |\mathcal{N}|))$\;
      $S_N \leftarrow \mathcal{N}.\text{dequeue}(\le k_N)$\;
      $S_F \leftarrow \text{select } (k - |S_N|) \text{ features from } \mathcal{F}$\;
      $S \leftarrow S_N \cup S_F$\;
            
      $G \leftarrow \text{\groupllm}.\text{complete}(S)$\;
      $c \leftarrow \text{\instantiatellm}.\text{instantiate}(G)$\; 
      
      Execute $c$ and measure coverage\;
      $C' \leftarrow \mathsf{Cov}.\text{sample}()$\;
      
      \If{$C' > C$}{
          $\mathcal{N} \leftarrow \mathcal{N} \cup (G \setminus S)$\;
          $C \leftarrow C'$\;
      }
  }
\end{algorithm}

Algorithm~\ref{alg:feature-selection} formalizes this feedback loop.
In each iteration, \tool constructs an initial feature set $S$ 
using a hybrid sampling strategy:
Here, $\text{UniformInt}(a,b)$ denotes uniform sampling of an integer from the inclusive range $[a,b]$, which determines how many novel features are drawn from $\mathcal{N}$ in the current iteration.
\textit{(1) Exploitation:} It dequeues up to $k$ ``novel'' features from $\mathcal{N}$, a high-priority queue containing features that have previously demonstrated coverage gains.
\textit{(2) Exploration:} It fills the remaining slots (up to size $k$) by sampling from the global pool $\mathcal{F}$ to maintain semantic diversity.

After $\text{GroupLLM}$ expands $S$ into the complete group $G$ and $\text{\instantiatellm}$ generates the program $c$, the fuzzer executes $c$ against the target compiler. If the coverage $C'$ exceeds the previous global state $C$, the newly introduced features $(G \setminus S)$ are deemed validated and promoted to the novel feature queue $\mathcal{N}$. This dynamic enrichment ensures that the fuzzer progressively biases its generation toward semantic combinations that successfully stress the compiler's optimization passes.

\section{Evaluation}
We evaluate \tool on the following research questions:

\begin{itemize}[]
    \item[\textbf{RQ1:}]  How does \tool compare against existing compiler fuzzers?
    \item[\textbf{RQ2:}] What is the quality of the extracted feature pool by \extractllm?
    \item[\textbf{RQ3:}] What is the effect of \groupllm on feature-group synthesis and the resulting programs?
    \item[\textbf{RQ4:}] Can \tool uncover new bugs in real-world and widely-used compilers?
\end{itemize}

\subsection{Implementation}
\label{sec:implementation}

To build a dataset of bug reports, bug-triggering programs, and fix histories, we mine resolved bugs from the GCC Bugzilla and the official GCC repository.
Each GCC bug report typically includes a bug-triggering program along with a unique bug ID, and developers often reduce these programs and commit them as regression tests.
Bug IDs allow us to trace their corresponding fix in the repository history. \looseness=-1

We collect 18,158 GCC bugs and use Qwen2.5-Max~\cite{alibabacloudqwenmodels} to extract 74,362 features, which are organized into 18,158 feature groups, each containing an average of 4 features.
We use Qwen3-4B~\cite{alibabacloudqwenmodels} as the base model for \groupllm and fine-tune it on these feature groups, yielding approximately 91MB training data.
Since program instantiation demands stronger code generation capabilities, we employ a larger model, i.e., Qwen3-32B~\cite{alibabacloudqwenmodels}, as \instantiatellm.

During the fuzzing process, to balance the diversity and coherence of feature groups, and given that each group typically contains around four features, we randomly sample two features as the initial input to \groupllm.
\tool uses AFL++~\cite{fioraldi2020afl++} to instrument the target compilers for coverage feedback. 
\tool is primarily implemented in Python.

\subsection{Evaluation Setups}

\topic{Answering RQ1.}
We evaluate \tool on two widely used compilers, GCC-13 and LLVM-18, and compare it against four representative compiler fuzzing approaches: mutation-based, rule-based, LLM-based, and function-synthesis methods.
From these approaches, we select representative fuzzers that are capable of generating C/C++ programs for testing GCC and LLVM.

\textit{Mutation-based fuzzers (\metamut and \mutfall)} generate new programs by applying random mutations to existing valid programs. 
We select two SOTA mutation-based fuzzers: 
\metamut~\cite{MetaMut2024} uses an LLM agent to design and implement high-quality mutators, while \mutfall~\cite{Mut4All2025} leverages an LLM agent to design more mutators based on bug reports.
Both \metamut and \mutfall use the official GCC test suites as seed programs, which include all bug-triggering programs used by \tool.
To examine whether increasing the number of mutators can improve the diversity of generated programs, we adapt \mutfall by integrating the coverage feedback mechanism from \metamut, enabling coverage-guided fuzzing.
\metamut applies its LLM-designed mutators under coverage guidance, whereas the default \mutfall does not use coverage feedback.

\textit{Rule-based fuzzers (YARPGen)} generate programs based on predefined rules and templates. 
We select \yarpgen~\cite{YARPGen2020}, a random program generator for C and C++. 
We exclude the classic \csmith~\cite{CsmithPLDI11} as it has reached apparent saturation and struggles to find bugs in recent versions of GCC and LLVM~\cite{YARPGen2020}.

\textit{LLM-based fuzzers (Fuzz4All and \bugft)} leverage LLMs to generate programs. 
We select \fuzzfall~\cite{fuzz4all2024}, which uses LLMs as both the generation and mutation engines. 
To the best of our knowledge, it represents SOTA in LLM-based fuzzing for LLVM and GCC.
We configure \fuzzfall to use the same Qwen3-32B model as \tool.
We also include a direct fine-tuning baseline, \bugft, in which Qwen3-32B is fine-tuned on the bug-triggering programs collected from our bug reports. 
\bugft evaluates the paradigm of encoding bug-prone semantics in model parameters without explicit feature extraction and recombination.

\textit{Function-synthesis fuzzers (LegoFuzz)} generate programs by interleaving functions that meet specific criteria. \creal~\cite{Creal2024} identifies and collects such functions.
Building on this, \legofuzz~\cite{LegoFuzz2025} extends the approach by leveraging LLMs to transform non-compliant functions to meet specific criteria, thereby increasing functional diversity. 
We select \legofuzz as the representative function-synthesis fuzzer for our evaluation.

\textbf{\tool Effectiveness (Coverage and Crashes):} 
The experiment evaluates both code coverage and unique crashes in GCC-13 and LLVM-18 over 24 hours of fuzzing.
Each fuzzer is executed five times on each compiler, resulting in a total of $2\ (\text{compilers}) \times 5\ (\text{runs}) \times 7\ (\text{fuzzers}) \times 24\ \text{CPU hours} = 1{,}680$ CPU hours.
Line coverage is measured using \texttt{gcov}~\cite{gcc-gcov} for both GCC and LLVM. 
All experiments are conducted on a machine equipped with an Intel Core i7-11700K CPU. For LLM-based fuzzers, \tool, \fuzzfall, and \bugft deploy their language models on an NVIDIA H800 GPU. 
The five parallel runs of each LLM-based fuzzer share the GPU simultaneously.

\noindent\topic{Answering RQ2.}
To evaluate the quality of the features extracted by \tool, we construct feature groups by randomly sampling features (FFRandom) without using \groupllm. 
This setup allows us to examine how the number of extracted features impacts code coverage and the number of unique crashes.

\noindent\topic{Answering RQ3.}
To evaluate the improvements introduced by \groupllm, we assess the coherence of feature groups by measuring their semantic redundancy and diameter. 
We further examine the validity of the resulting programs through the compilation pass rate. 
Finally, we investigate \groupllm's ability to discover novel features by comparing its coverage distribution to that of FFRandom.

\textbf{Feature Group Coherence (Semantic Redundancy and Diameter):} 
A coherent feature group should consist of features that are related within a shared program context while capturing a range of distinct semantics, rather than repeating similar ones.
To quantify coherence, we employ semantic metrics: semantic redundancy and semantic diameter.
We use SBERT~\cite{sbert2019} to compute embeddings that represent the semantics of individual features.
Both metrics are calculated based on these embeddings.
Semantic redundancy measures the degree of semantic overlap (i.e., shared program context) among features within a group,
whereas semantic diameter captures the extent of semantic diversity (i.e., distinct semantics).
We formally define both metrics as follows.

\paragraph{Semantic Redundancy}
Given a feature group $G = \{f_1, \dots, f_n\}$, we embed each feature into a vector
$\mathbf{e}_i \in \mathbb{R}^d$ using SBERT.
Semantic redundancy is defined as the average pairwise cosine similarity among feature embeddings,
excluding self-pairs and near-duplicate pairs:
\[
\text{Redundancy}(G)
=
\frac{1}{|\mathcal{P}|}
\sum_{(i,j)\in\mathcal{P}}
\cos(\mathbf{e}_i, \mathbf{e}_j),
\]
where
\[
\mathcal{P}
=
\left\{
(i,j)\ \middle|\ 
i \neq j \ \wedge\ \cos(\mathbf{e}_i, \mathbf{e}_j) < \tau
\right\},
\]
and $\tau(0.95)$ is a similarity threshold used to filter out near-duplicate feature pairs.

\paragraph{Semantic Diameter}
Semantic diameter captures the maximum semantic span within a feature group and is defined as
\[
\text{Diameter}(G)
=
\max_{i \neq j}
\left( 1 - \cos(\mathbf{e}_i, \mathbf{e}_j) \right).
\]

\noindent\topic{Answering RQ4.}
We use \tool to test the latest stable versions of GCC-15 and LLVM-21 over a three-day period, running on a single NVIDIA H800 GPU and an Intel Core i7-11700K CPU. 
The discovered bugs have been reported to the GCC and LLVM communities, and we track their confirmation and resolution status.

\subsection{RQ1: Comparison with Existing Compiler Fuzzers}

\begin{figure*}[h]
  \centering
  \captionsetup{skip=2pt}
  \captionsetup[sub]{skip=1pt}

  \begin{subfigure}{0.496\textwidth}
    \includegraphics[width=\linewidth]{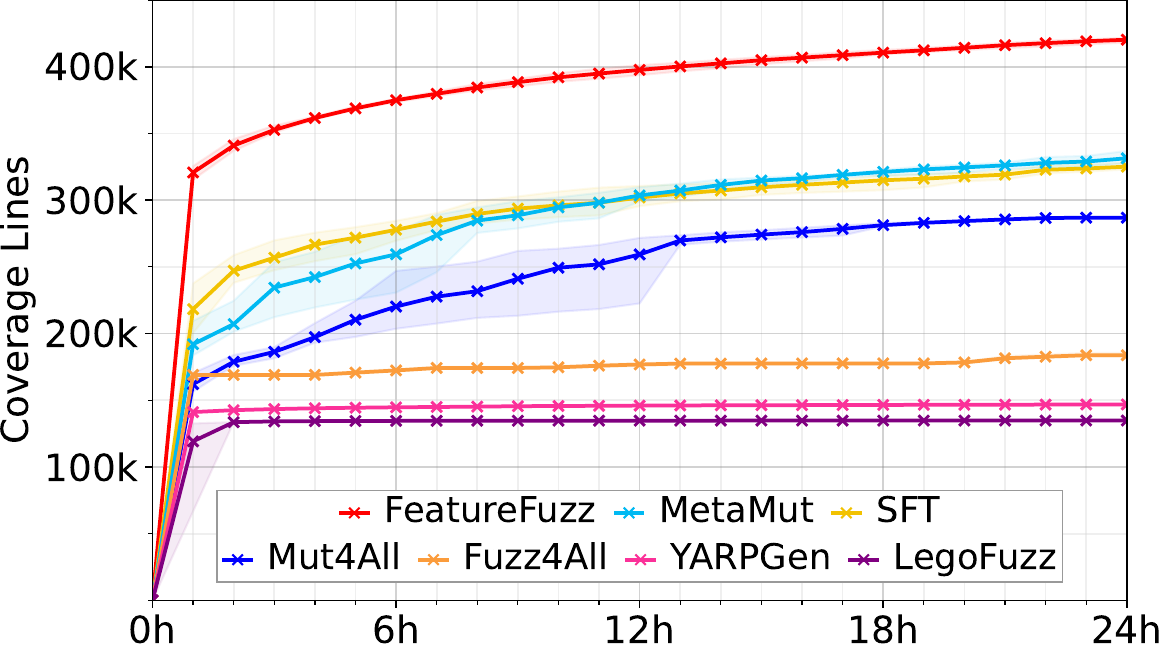}
    \caption{GCC}
    \label{fig:gcc_line_cov}
  \end{subfigure}
  \begin{subfigure}{0.496\textwidth}
    \includegraphics[width=\linewidth]{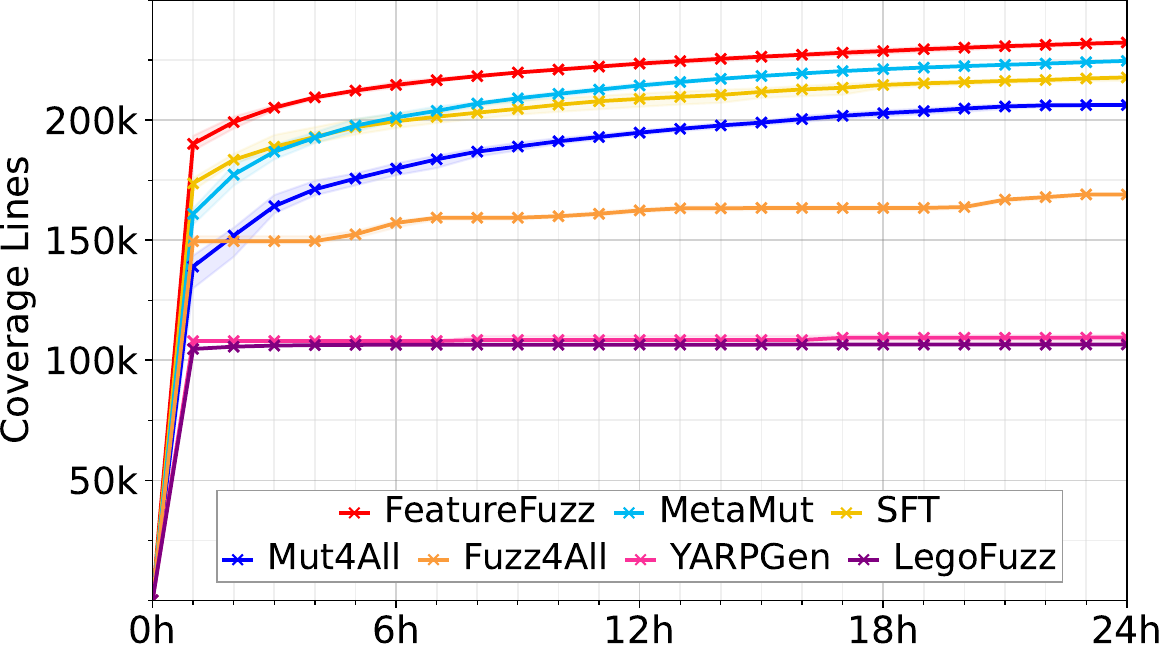}
    \caption{LLVM}
    \label{fig:llvm_line_cov}
  \end{subfigure}
  \caption{Line coverage trends of each fuzzer for GCC-13 and LLVM-18. \tool outperforms existing fuzzers.}
  \label{fig:line_coverage_trends}
\end{figure*}

\subsubsection{Code Coverage and Validity}
\label{sec:coverage_results}
Figure~\ref{fig:line_coverage_trends} presents the line coverage trends of different fuzzers on GCC-13 and LLVM-18 over a 24-hour fuzzing period. Solid lines indicate the average coverage across five runs, and shaded regions represent the range between the minimum and maximum coverage. \metamut and \mutfall exhibit fluctuations, but ultimately reach a stable state.

Compared to other fuzzers, \tool achieves higher coverage early and continues to maintain superior coverage growth over time (Figure~\ref{fig:line_coverage_trends}).
After a 24-hour fuzzing campaign, \tool achieves the highest line coverage among all evaluated fuzzers, reaching 420.24K lines on GCC and 232.30K lines on LLVM (Table~\ref{tab:coverage_validity_all_fuzzers}). 
\tool improves line coverage by 26.8\% on GCC and 3.4\% on LLVM over the strongest baseline, \metamut.

\bugft and \mutfall both learn from historical bugs, but rank only third and fourth. This suggests that directly encoding bug-prone knowledge into model weights or mutators is helpful, but still insufficient for systematically reusing and recombining the underlying semantics. \metamut ranks second and remains the strongest baseline. Other tools, including the LLM-based fuzzer \fuzzfall, the rule-based fuzzer \yarpgen, and the function-synthesis fuzzer \legofuzz, yield lower coverage results.

The efficiency of \tool stems from its ability to generate high-quality programs. Mutation-based fuzzers generate nearly 350K programs within 24 hours, yet only 10K programs (approximately 2.9\%) increase coverage in GCC and 9.5K programs (approximately 2.7
\%) in LLVM. 
In contrast, although \tool achieves higher coverage in the early stages, \tool generates approximately 50K programs but achieves 9.5K coverage-increasing programs(19\%) in GCC and 8K coverage-increasing programs (16\%) in LLVM. 
While \tool maintains a lower throughput than mutation-based approaches, its higher ratio of coverage-increasing programs demonstrates the ability of \tool to sustainably explore compiler behaviors more effectively.

\tool attains 39.76\% validity, ranking sixth among the evaluated fuzzers (Table~\ref{tab:coverage_validity_all_fuzzers}). Combined with the throughput results above, this shows that \tool does not achieve high coverage by generating more programs or by achieving higher validity. Instead, its advantage comes from its stronger ability to preserve and recombine bug-triggering semantics.

\begin{table}[h!]
\centering
\setlength{\tabcolsep}{5pt}
\renewcommand{\arraystretch}{1.12}
\caption{Program validity and coverage of different fuzzers. }
\label{tab:coverage_validity_all_fuzzers}
\begin{tabular}{l c l}
\toprule
\textbf{Fuzzer} & \textbf{Valid(\%)} & \textbf{Coverage (GCC / LLVM)} \\
\midrule
\rowcolor{blue!12}\tool & 39.76\% & \textbf{420,239 / 232,304} \\
\metamut  & 84.52\% & 331,381 / 224,654 \\
\bugft  & 72.89\% & 325,110 / 217,780 \\
\mutfall  & 47.92\% & 286,987 / 206,242 \\
\fuzzfall  & 35.67\% & 183,780 / 169,059 \\
\yarpgen & 79.03\% & 146,800 / 109,423 \\
\legofuzz & 99.39\% & 134,825 / 106,570 \\
\bottomrule
\end{tabular}
\end{table}

\subsubsection{Coverage Distributions}

\begin{figure}[h!]
    \centering
    \begin{subfigure}{\linewidth}
        \includegraphics[width=\linewidth]{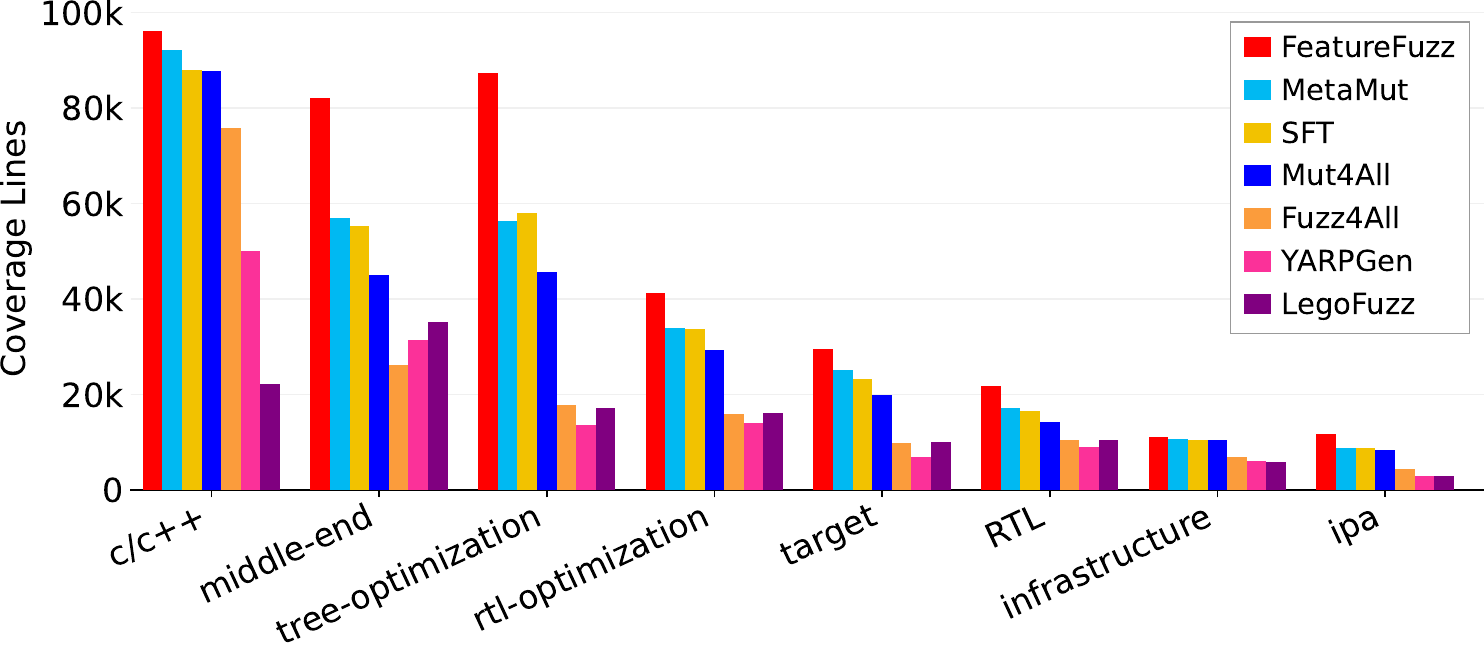}
        \caption{GCC}
        \label{fig:dist_gcc}
    \end{subfigure}
    
    \begin{subfigure}{\linewidth}
        \includegraphics[width=\linewidth]{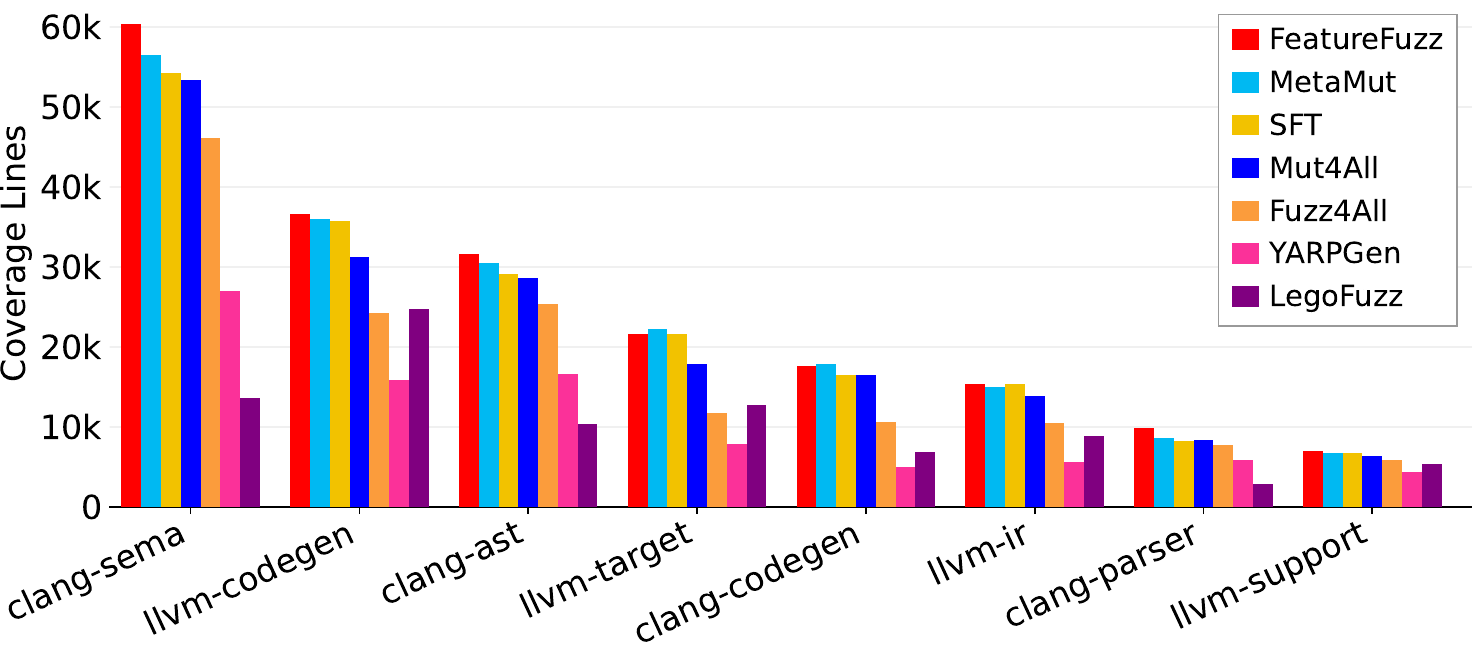}
        \caption{LLVM}
        \label{fig:dist_llvm}
    \end{subfigure}
    
    \caption{Line coverage across the top 8 components in GCC and LLVM.}
    \label{fig:coverage_distribution}
\end{figure}

We further analyze the distribution of coverage across compiler components to identify the sources of the observed coverage gains.
Figure~\ref{fig:coverage_distribution} presents the line coverage across the top 8 compiler components of GCC-13 and LLVM-18 after 24 hours of fuzzing.

For GCC, \tool achieves the highest coverage across the front-end (C/C++), middle-end, optimizations (tree-optimization, rtl-optimization), and back-end(target). 
A similar trend is observed on LLVM. 
Although the overall coverage gains achieved by \tool on LLVM are more modest than those on GCC, \tool still achieves the highest coverage in many components, including the front-end (clang-sema and clang-ast), the middle-end (llvm-ir) and back-end(llvm-codegen). 
\fuzzfall focuses primarily on the C/C++ front-end. 
This suggests that, although \tool also generates programs based on LLMs like \fuzzfall and \bugft, it benefits from features related to compiler optimizations and even the back-end.
As a result \tool explores the compiler in further depth.
The line coverage achieved by \tool comes from multiple components of GCC and LLVM, rather than focusing on just a few specific components.

\subsubsection{Unique Crashes}

To evaluate and compare the bug-finding capabilities of different fuzzers, we collect crashes during experiments.
Crash uniqueness is determined based on the type of failure. Assertion failure crashes are identified by error messages, while other crashes, such as segmentation faults and out-of-memory errors, are distinguished through stack trace analysis~\cite{MetaMut2024, Mut4All2025}.




Table~\ref{tab:upsert_bugs} summarizes the 243 unique crashes found by the evaluated fuzzers, including both fuzzer-specific and overlapping bugs.
\tool outperforms \metamut, the best result among \metamut, \bugft, \mutfall, \fuzzfall, \yarpgen, and \legofuzz, detecting 2.78 times as many.
Among all crashes, \tool identifies 167 crashes, \metamut finds 60 crashes, \bugft discovers 48 crashes, \mutfall discovers 11 crashes, and \fuzzfall uncovers 12 crashes,
while \yarpgen and \legofuzz do not detect any crashes.
\tool detects the majority of crashes, accounting for 68.72\% (167 out of 243) of all crashes.

\newcommand{\hit}{\raisebox{0.05ex}{\normalsize\ding{51}}}

\begin{table}[t]
\caption{Intersection crashes detected by each fuzzer. \legofuzz and \yarpgen are excluded because they found no crashes.}
\label{tab:upsert_bugs}
\centering
\small
\setlength{\tabcolsep}{5pt}
\renewcommand{\arraystretch}{1.08}
\begin{threeparttable}
\rowcolors{2}{white}{blue!6}
\begin{tabular}{@{}lccccc@{}}
\toprule
\rowcolor{white}
\textbf{Crashes} & \textbf{\tool} & \textbf{\metamut} & \textbf{\bugft} & \textbf{\fuzzfall} & \textbf{\mutfall} \\
\midrule
\textbf{131} & \hit &  &  &  &  \\
\textbf{38}  &  & \hit &  &  &  \\
\textbf{23}  &  &  & \hit &  &  \\
\textbf{12}  & \hit &  & \hit &  &  \\
\textbf{10}  & \hit & \hit &  &  &  \\
\textbf{5}   & \hit &  &  & \hit &  \\
\textbf{5}   &  &  &  &  & \hit \\
\textbf{4}   &  &  &  & \hit &  \\
\textbf{3}   & \hit & \hit & \hit &  & \hit \\
\textbf{3}   & \hit & \hit & \hit &  &  \\
\textbf{3}   & \hit &  & \hit & \hit &  \\
\textbf{3}   &  & \hit & \hit &  &  \\
\textbf{2}   &  & \hit &  &  & \hit \\
\textbf{1}   &  & \hit & \hit &  & \hit \\

\hiderowcolors
\midrule
\textbf{Total} & \textbf{167} & \textbf{60} & \textbf{48} & \textbf{12} & \textbf{11} \\
\bottomrule
\end{tabular}
\begin{tablenotes}
\footnotesize
\item  Crashes denotes the number of unique bugs revealed by the corresponding fuzzer intersection.
\end{tablenotes}
\end{threeparttable}
\end{table}

Other fuzzers exhibit different bug-finding capabilities.
\metamut and \mutfall together trigger 65 unique crashes, with \metamut significantly outperforming \mutfall (60 vs. 11).
\bugft discovers 48 crashes, outperforming \mutfall and \fuzzfall but still remaining well behind \tool and \metamut.
Despite achieving lower coverage compared to \mutfall, \fuzzfall still discovers 12 unique crashes(12 vs. 11).
We observe that \fuzzfall generates many structurally similar programs, suggesting that these patterns can effectively expose compiler-specific defects.
Both \yarpgen and \legofuzz detect no crashes. 
These tools are primarily designed to explore aggressive optimizations, rather than focusing on general compiler crashes.

\tool identifies 167 crashes, 131 of which are found exclusively by our approach. Only 21.56\% of its crashes are shared with existing fuzzers (Table~\ref{tab:upsert_bugs}), showing that \tool effectively explores deep compiler behaviors that remain inaccessible to other tools.

\subsubsection{Crash Distribution Across Compiler Components}

\begin{figure}[h]
    \centering
    \begin{subfigure}{\linewidth}
        \includegraphics[width=\linewidth]{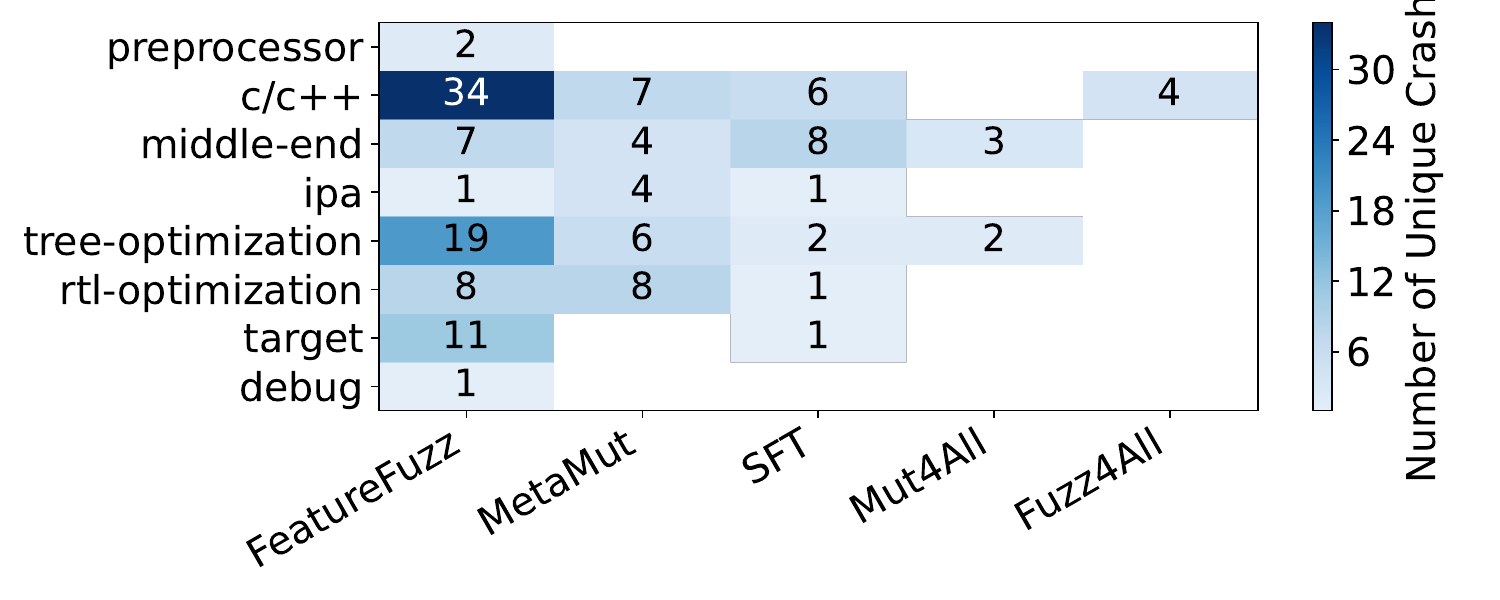}
        \caption{GCC}
        \label{fig:crash_heatmap_gcc}
    \end{subfigure}

    \begin{subfigure}{\linewidth}
        \includegraphics[width=\linewidth]{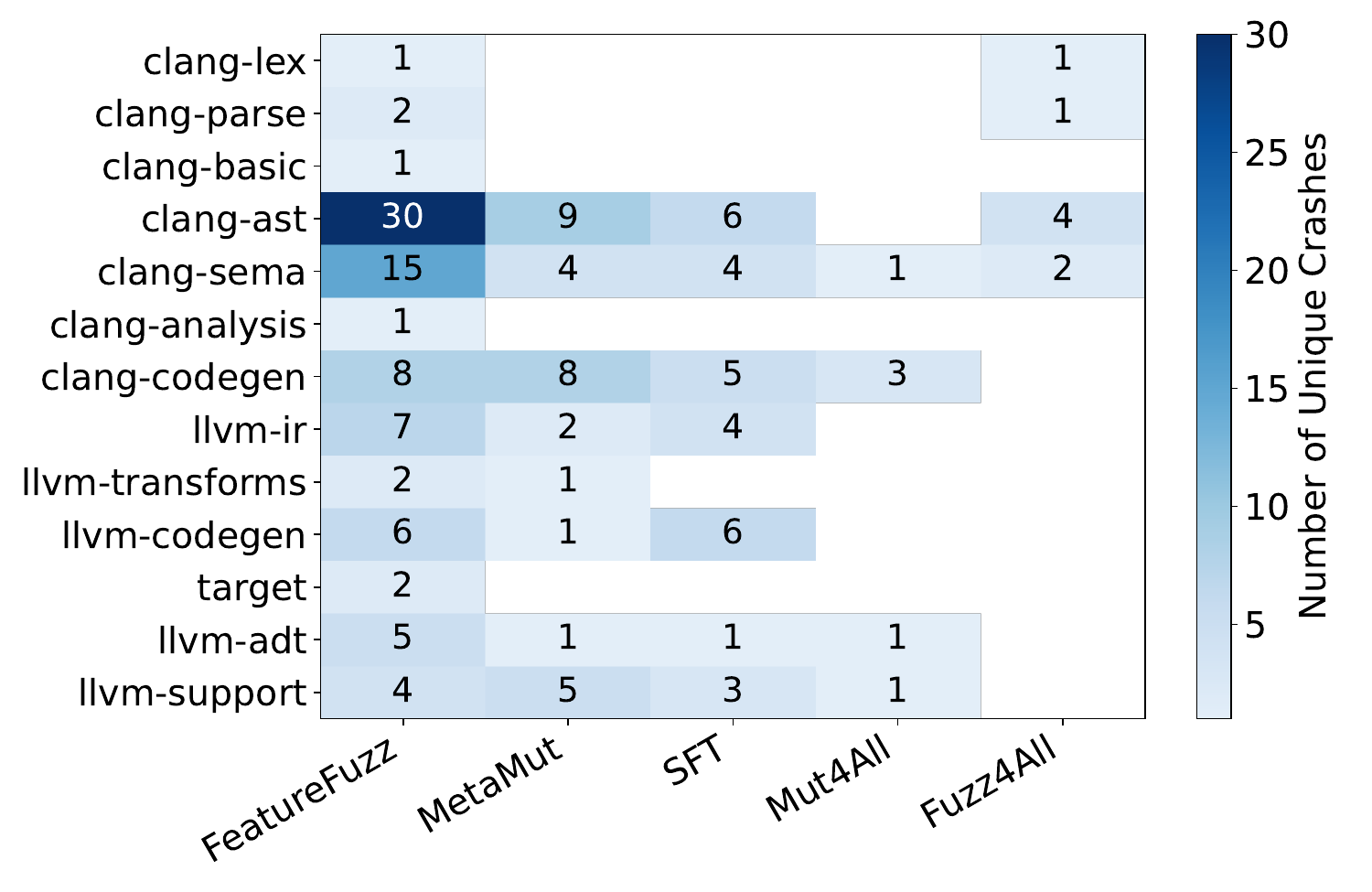}
        \caption{LLVM}
        \label{fig:crash_heatmap_llvm}
    \end{subfigure}
    \caption{Unique crashes across different components in GCC and LLVM. Crashes detected by 
                \tool cover all components.}
    \label{fig:crashes_heatmap}
\end{figure}

Figure~\ref{fig:crash_heatmap_gcc} and Figure~\ref{fig:crash_heatmap_llvm} show the distribution of crashes across different components in GCC and LLVM, respectively.
The order of components from top to bottom reflects the compilation pipeline—from front-end to back-end.
The \textit{target} component in both GCC and LLVM refers to the
architecture-specific back end responsible for platform-specific code
generation and optimization. The \textit{debug} component in GCC spans the
compiler pipeline, while \textit{llvm-adt} and \textit{llvm-support} provide
foundational data structures and utilities for LLVM. These target and auxiliary
components appear near the bottom of the corresponding heatmaps.

For both GCC and LLVM, unique crashes discovered by \tool are widely distributed across the entire compilation pipeline rather than being confined to a single component.
For GCC, \tool discovers 83 crashes across the compilation pipeline, including
36 in front-end components such as the preprocessor and C/C++ front end, 35 in
middle-end and optimization components such as IPA, tree optimization, and RTL
optimization, and 11 in the target back end, plus one in the debug component.
For LLVM, \tool discovers 84 crashes, including 50 in front-end components such
as Clang AST and Sema, 17 in IR-generation and optimization components such as
Clang CodeGen, LLVM IR, and LLVM Transforms, and 8 in back-end components such
as LLVM CodeGen and Target, plus 9 in LLVM ADT and Support.
Compared to other fuzzers, \tool identifies bugs across a wider range of compiler components.
Specifically, \tool triggers crashes across 8 distinct GCC components and 13 LLVM components. 
In contrast, the breadth of other fuzzers is notably more restricted: \metamut affects 5 GCC and 8 LLVM components, \bugft reaches 6 GCC and 7 LLVM components, and \mutfall is limited to 2 and 4, respectively. Furthermore, \fuzzfall exhibits a narrow focus on the front-end, reaching only 1 component in GCC and 4 in LLVM.

\tool excels at deep testing of the compiler back-end, effectively reaching corner cases and components that are frequently overlooked by existing fuzzers.
\tool successfully triggered 11 unique crashes in GCC’s \texttt{target} component and two unique crashes within the LLVM \texttt{target} component. 
In contrast, competing fuzzers such as \metamut, \mutfall, and \fuzzfall failed to uncover any bugs in these critical components.

\vspace{-2em}
\subsubsection{LLM Cost}

\begin{table}[h]
\centering
\small
\setlength{\tabcolsep}{3pt}
\caption{LLM token usage and cost, estimated from public API pricing with OpenRouter as the reference provider.}
\label{tab:llm_cost}
\begin{tabular}{lcccc}
\toprule
\textbf{Tool} & \makecell{\textbf{Group}\\\textbf{LLM}} & \makecell{\textbf{Input tokens}\\\textbf{per program}} & \makecell{\textbf{Output tokens}\\\textbf{per program}} & \makecell{\textbf{Cost}\\\textbf{(1K programs)}} \\
\midrule
\tool & on & 549 & 420 & \$0.145 \\
\tool & off & 343 & 304 & \$0.100 \\
\fuzzfall & -- & 521 & 285 & \$0.110 \\
\bugft & -- & 39 & 148 & \$0.039 \\
\bottomrule
\end{tabular}
\footnotesize
Other evaluated fuzzers do not invoke LLMs during generation and are therefore omitted from this comparison.
\end{table}

Table~\ref{tab:llm_cost} shows that \tool maintains a practical LLM cost comparable to prior LLM-based fuzzers such as \fuzzfall, even with \groupllm enabled. 
\tool uses the small Qwen3-4B model for \groupllm in practice.
Since public API pricing for Qwen3-4B is unavailable, we estimate the \groupllm cost using Qwen3-32B pricing. 
This gives a conservative upper bound, and the actual cost is lower. 
\bugft incurs low cost because fine-tuning allows simple prompts without format constraints or detailed instructions, and its generated programs are also short, as the bug-triggering programs in its training data are generally short.

\noindent\textbf{Answer to RQ1.}
\tool outperforms all baselines in coverage breadth and depth and in defect
number and diversity, finding 2.78$\times$ as many unique crashes as the
second-best fuzzer.

\subsection{RQ2: Quality of Extracted Feature Pool}

%

\begin{table}[h]
\centering
\small
\caption{Unique crashes and union coverage of FFRandom under different feature pool sizes over 72 hours (3 runs).}
\label{tab:random_pool_scaling}
\begin{tabular}{lc @{\hskip 10pt} c @{\hskip 10pt} c @{\hskip 2pt} c}
\hline
\multirow{2}{*}{\textbf{Pool Size}} 
& \multirow{2}{*}{\textbf{Crashes}} 
& \multicolumn{2}{c}{\textbf{Union Coverage}} \\
\cline{3-4}
&  & \textbf{GCC} & \textbf{LLVM} \\
\hline
1/1 & 74 & 429,749 (Avg.400,886)  &  243,894 (Avg.232,964)\\
1/2 & 82 & 428,064 (Avg.395,967) & 243,705 (Avg.231,389) \\
1/4 & 84 & 428,330 (Avg.399,506) & 243,467 (Avg.231,772) \\
1/8 & 74 & 423,864 (Avg.397,344) & 241,693 (Avg.231,386) \\
\hline
\end{tabular}
\end{table}

\begin{figure}[h]
    \centering
    \includegraphics[width=\linewidth]{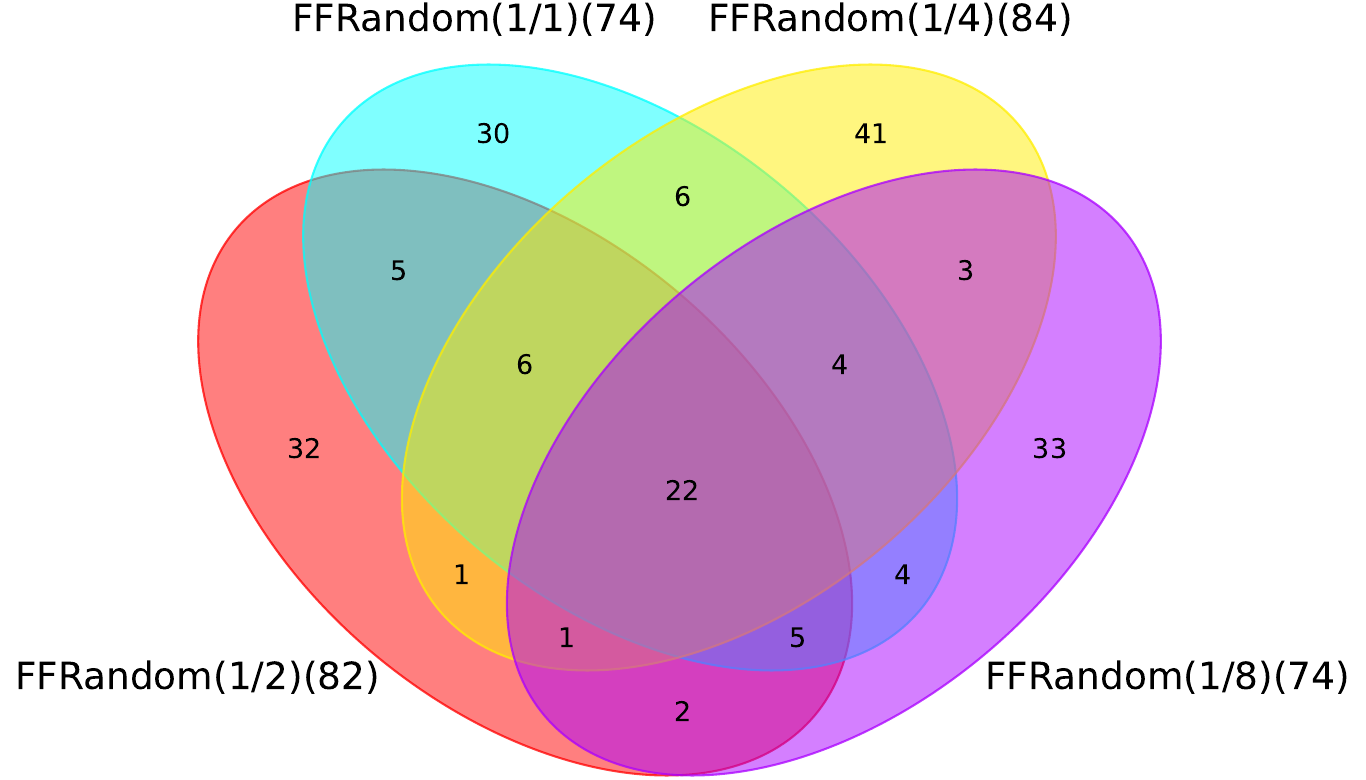}
    \caption{Venn diagram of unique crashes discovered by FFRandom with different pool sizes over 72 hours (3 runs).}
    \label{fig:venn_bugs_random_pool_size}
\end{figure}

To evaluate the quality of the extracted features, we use random sampling (FFRandom) to create feature groups and generate programs under varying feature pool sizes.
We evaluate FFRandom in terms of unique crashes and code coverage.

Specifically, we randomly subsample the collected feature pool to 100\%, 50\%, 25\%, and 12.5\% of its size. 
For each pool size, we conduct three fuzzing runs and test the generated programs on GCC-13 and LLVM-18, with each run lasting 24 hours.
Table~\ref{tab:random_pool_scaling} summarizes the number of unique crashes (across 3 runs) and the union line coverage from the three runs.

FFRandom(1/8) produces an average of 397,344 lines on GCC and 231,386 lines on LLVM, surpassing the best baseline, \metamut, which achieves 331,381 lines on GCC and 224,654 lines on LLVM (Figure~\ref{fig:line_coverage_trends}).
The FFRandom(1/8) configuration (3 runs, 72 hours total) discovers 74 crashes, surpassing the 60 crashes found by programs generated by \metamut across 5 GCC runs and 5 LLVM runs (240 hours total).
The increased coverage and number of crashes indicate that the features extracted by \extractllm are intrinsically effective. Even without feature group completion, randomly sampling a small subset of these features enables more extensive compiler exploration than existing fuzzers.

As the feature pool size increases, the union coverage generally improves in both LLVM and GCC.
However, the number of unique crashes and the average coverage do not show a consistent upward trend (Table~\ref{tab:random_pool_scaling}).
FFRandom(1/2) detects 82 unique crashes, while FFRandom(1/1) detects fewer, with 74 unique crashes.
This indicates that the features still have significant exploration potential, and 24 hours of fuzzing may not be sufficient to fully explore this expanded space.
Figure~\ref{fig:venn_bugs_random_pool_size} also supports this observation.
Even when combining the bugs found across 3 runs, different pool sizes still reveal a large number of unique bugs. 
For example, FFRandom(1/4) alone accounts for 41 such unique bugs.

\noindent\textbf{Answer to RQ2.}
Features extracted by \extractllm provide a broad exploration space: even
random sampling from one eighth of the pool outperforms all baselines in both
coverage and unique crashes.

\subsection{RQ3: Ablation Study and Novelty of Features Induced by \groupllm}

\begin{table}[h]
\centering
\small
\caption{Ablation results of \tool variants over three GCC and three LLVM runs.}
\label{tab:recombination_ablation}
\begin{tabular}{lccc}
\toprule
\textbf{Variant} & \textbf{Coverage(GCC/LLVM)} & \textbf{Crashes} & \textbf{Valid(\%)}\\
\midrule
\tool & 420,239 / 232,304 & 129 & 39.76 \\
\toolft {\scriptsize (w/o feedback)} & 412,745 / 230,646 & 115 & 42.72\\
\toolrandomcov {\scriptsize (w/o GroupLLM)} & 403,375 / 232,733 & 127 & 24.41\\
\toolrandom {\scriptsize (w/o both)} & 400,886 / 232,964 & 127 & 23.60 \\
\toolsingle {\scriptsize (single feat.)} & 361,508 / 218,120 & 51 & 73.75\\
\bottomrule
\end{tabular}
\end{table}


Table~\ref{tab:recombination_ablation} isolates \groupllm, coverage feedback, and
multi-feature recombination. The full \tool enables all three; \toolft removes
feedback; \toolrandomcov removes \groupllm; \toolrandom removes both; and
\toolsingle uses one feature at a time. \toolsingle has the lowest coverage and
fewest crashes, confirming the value of recombination. \tool and \toolft are substantially more valid than the random-grouping variants, indicating that \groupllm improves program validity. 
However, \toolft favors \groupllm's training distribution and slightly reduces LLVM coverage and crashes. 
The full tool restores exploration via coverage feedback while preserving GroupLLM's validity gains.

\begin{table}[h]
\centering
\caption{
    \emph{Semantic redundancy and diameter} of feature groups under different construction strategies.
    Lower diameter and moderate redundancy indicate more cohesive yet non-redundant feature groups.
}
\label{tab:semantic_stats_transposed}
\begin{tabular*}{\linewidth}{@{\extracolsep{\fill}}lccc}
\hline
\textbf{Variant} & \textbf{Redundancy} & \textbf{Diameter} & \textbf{CrashOnValid} \\
\hline
CFG$^{*}$ & $0.640 \pm 0.112$ & $0.477 \pm 0.153$ & 91.72\%$^{*}$ \\
\tool & $0.619 \pm 0.070$ & $0.569 \pm 0.114$ & 21.53\% \\
\toolft & $0.611 \pm 0.069$ & $0.592 \pm 0.121$ & 23.46\% \\
\toolrandomcov & $0.466 \pm 0.093$ & $0.638 \pm 0.143 $ & 11.81\% \\
\toolrandom & $0.468 \pm 0.092$ & $0.638 \pm 0.143$  & 11.42\% \\
\hline
\end{tabular*}
\footnotesize
$^{*}$CFG denotes \emph{Collected Feature Groups} extracted by \extractllm. 
For CFG, \emph{CrashOnValid} represents the fraction of valid programs among the crash bugs.
\end{table}
\groupllm improves validity by repairing dependencies among features that are ignored by \toolrandom. To explain this effect, we first analyze the semantic properties of groups constructed under different strategies and then the validity of the resulting programs.

A coherent feature group should contain related features rather than isolated ones, while still preserving some diversity.
We therefore use the collected feature groups (CFG) from \extractllm as a reference set and measure semantic redundancy and semantic diameter. 
For \tool, \toolft, \toolrandomcov, and \toolrandom, we randomly sample 50K feature groups from each method and compute the two metrics for comparison.

Table~\ref{tab:semantic_stats_transposed} reports the semantic redundancy and semantic diameter of feature groups constructed under different strategies.
CFG exhibits the highest redundancy (0.640) and the lowest diameter (0.477).
Both \tool and \toolft use \groupllm to synthesize feature groups. 
They exhibit lower redundancy than CFG but significantly higher redundancy than \toolrandomcov and \toolrandom. In contrast, for diameter, their values are higher than those of CFG but considerably lower than those of \toolrandomcov and \toolrandom.
These results indicate that feature groups synthesized by \groupllm achieve more reasonable redundancy and diameter than random sampling, making them closer to the CFG where features from the same program are naturally more coherent.
Such groups are easier for \instantiatellm to realize as valid programs.
Appendix~\ref{app:semantic-correlation} further analyzes how these metrics relate to compilation success.

To understand whether \groupllm contributes to new exploration through newly discovered features, we analyze the coverage overlap across different variants of \tool.
We compare the union coverage of \tool, \toolft, and \toolrandom, each run three times per compiler (GCC and LLVM).
For each pair of tools, we measure:
(i) the overlapping coverage regions,
(ii) the coverage regions unique to each tool, and
(iii) the Jaccard between their union coverage. 
\toolrandom was run 6 times on both GCC and LLVM, and we compare the coverage between the initial three ($R_{1\text{--}3}$) and subsequent three runs ($R_{4\text{--}6}$).


\begin{table}[h]
\centering
\small
\caption{Coverage differences between \tool, \toolft, and \toolrandom. \toolrandom uses all the collected features. Jaccard is the ratio of the overlapping coverage to the total coverage of both tools. }
\begin{tabular}{lccc@{\hspace{-0.0em}}c}
\toprule
\small{\textbf{Variant}} & \small{\textbf{Overlap}} & \small{\textbf{Unique}} & \small{\textbf{Jaccard(\%)}} \\ \midrule
\rowcolor{gray!15} \multicolumn{4}{l}{GCC} \\
\tool\ vs \toolrandom & 420,455 & 31,663 vs 13,442 & 90.31\% \\
\toolft\ vs \toolrandom & 417,312  & 27,095 vs 16,585 & 90.52\% \\ 
\toolrandom($R_{1\text{--}3}$) vs \toolrandom($R_{4\text{--}6}$) & 419,692 & 14,205 vs 13,006 & 93.91\% \\
\midrule
\rowcolor{gray!15} \multicolumn{4}{l}{LLVM} \\ 
\tool\ vs \toolrandom & 236,733 & 7,796 vs 7,919  & 93.78\% \\
\toolft\ vs \toolrandom & 234,990 & 6,346 vs 9,662   & 93.62\% \\
\toolrandom($R_{1\text{--}3}$) vs \toolrandom($R_{4\text{--}6}$) & 239,321 & 5,331 vs 4,876   & 95.91\% \\
\bottomrule
\end{tabular}
\label{tab:coverage_overlap}
\end{table}

All pairs share most coverage (Jaccard 0.90--0.95), as expected from their
common initial feature pool. 
\tool and \toolft explore coverage regions distinct from those of \toolrandom, with \tool uniquely covering 31,663 lines in GCC and 7,796 in LLVM (\toolft: 27,095 and 6,346, respectively). 
Their Jaccard similarities with \toolrandom are lower than those between two subsets of \toolrandom's own runs. 
A control experiment comparing the first three \toolrandom runs ($R_{1\text{--}3}$) with the last three ($R_{4\text{--}6}$) yields much higher similarity, 
confirming that the differences arise from new features rather than fuzzing randomness.
\noindent\textbf{Answer to RQ3.}
Overall, GroupLLM-enabled \tool achieves 68\% higher compilation success than
random grouping without GroupLLM or coverage feedback and adds 31,663 covered
lines on GCC and 7,796 on LLVM through newly introduced features.

\subsection{RQ4: Bug Finding}

\begin{table}[h]
\caption{Overview of the reported compiler bugs during the 72-hour fuzzing campaign.}
\label{tab:bug_finding}
\centering
\begin{tabular}{l|c|c|c}
\toprule & \textbf{GCC} & \textbf{LLVM} & \textbf{Total} \\ \midrule
\textbf{Reported} & \numberReportedBugGCC  & \numberReportedBugLLVM & \numberReportedBug \\ 
\rowcolor{gray!15} \multicolumn{4}{c}{Numbers of Reported Compiler bugs} \\ 
\textbf{Confirmed} & \numberConfirmBugGCC & \numberConfirmBugLLVM & \numberConfirmBug \\
\textbf{Assigned} & \numberAssignedBugGCC & \numberAssignedBugLLVM & \numberAssignedBug \\
\textbf{Duplicate} & \numberDuplicateBugGCC & \numberDuplicateBugLLVM & \numberDuplicateBug \\ 
\rowcolor{gray!15} \multicolumn{4}{c}{Numbers of Affected Compiler Components} \\ 
\textbf{Front-End} & \numberFrontEndBugGCC & \numberFrontEndBugLLVM & \numberFrontEndBug \\
\textbf{Middle-End} & \numberMiddleEndBugGCC & \numberMiddleEndBugLLVM & \numberMiddleEndBug \\
\textbf{Back-End} & \numberBackEndBugGCC & \numberBackEndBugLLVM & \numberBackEndBug \\
\rowcolor{gray!15} \multicolumn{4}{c}{Bug Breakdown} \\
\multicolumn{3}{l|}{\textbf{Assertion-Build Bugs}} & \numberNonReleaseBuildBug \\
\multicolumn{3}{l|}{\textbf{Release-Build Bugs}} & \numberReleaseBuildBug \\
\multicolumn{3}{l|}{\textbf{Release-Build Assertion Failures}} & \numberReleaseAssertionBug \\
\multicolumn{3}{l|}{\textbf{Release-Build Runtime Faults}} & \numberReleaseRuntimeFaultBug \\
\multicolumn{3}{l|}{\textbf{Release-Build Memory-Safety Bugs}} & \numberReleaseMemorySafetyBug \\
\multicolumn{3}{l|}{\textbf{Release-Build FPE}} & \numberReleaseFPEBug \\
\hline
\end{tabular}
\par\smallskip
{\footnotesize
\emph{Release-Build Runtime Faults} include failures such as segmentation faults and stack overflows; 
\emph{Release-Build Memory-Safety Bugs} stem from memory corruption; and 
\emph{Release-Build FPE} denotes floating-point exceptions.
\par}
\end{table}

Our 72-hour GCC and LLVM campaign reports \numberReportedBug bugs
(\numberReportedBugGCC GCC and \numberReportedBugLLVM LLVM), including
\numberValidBug triggered by compilable programs that better reflect realistic inputs. 
Developers confirmed \numberConfirmBug as zero-days and assigned \numberAssignedBug for fixing; only \numberDuplicateBug are duplicates, while the rest await triage. The bugs span the front end (\numberFrontEndBug), middle
end (\numberMiddleEndBug), and back end (\numberBackEndBug), demonstrating coverage of the compilation pipeline.

In light of the CI/CD threat model in Section~\ref{sec:threat_model}, these bugs show that \tool can uncover compiler defects with non-trivial security implications. The \numberReleaseBuildBug release-build bugs provide the strongest evidence, as they manifest in production binaries; they comprise memory-safety bugs (\numberReleaseMemorySafetyBug), runtime faults (\numberReleaseRuntimeFaultBug), FPEs (\numberReleaseFPEBug), and assertion failures (\numberReleaseAssertionBug). Among them, memory-safety bugs are the most severe due to memory corruption, while runtime faults can disrupt CI/CD runners by exhausting resources. FPEs and assertion failures are weaker signals but still break release compilers. In contrast, \numberNonReleaseBuildBug assertion-build bugs offer weaker yet meaningful warnings, revealing latent defects that may later surface in release builds.

\textbf{Heap Use-After-Free:} One representative case from the memory-safety bugs is a release-build GCC-15 defect that manifests as a heap use-after-free during register allocation. 
Listing~\ref{lst:gcc_uaf_case} shows the triggering program, which reproduces the bug when compiled with \texttt{g++ -O2} and diagnosed under AddressSanitizer instrumentation.

\vspace{6pt}
\begin{lstlisting}[language=C++,
caption={GCC-124711: GCC heap use-after-free during compilation.},
label=lst:gcc_uaf_case,
basicstyle=\footnotesize\ttfamily,
columns=flexible
]
__int128 large_operation() {
    __int128 result = 0;
    asm volatile (
        "vpxor %%xmm0, %%xmm0, %%xmm0\n"
        "vmovdqa (%[src]), %%xmm0\n"
        : [dst] "=Yz"(result)
        : [src] "r"(&result)
        : "%ymm0"
    );
    return result;
}
\end{lstlisting}

This bug is triggered by two features:
(1) using a function with a local \lstinline|__int128| value, and
(2) using \lstinline|asm volatile| with XMM registers and a memory operand tied to a variable.
To realize both features at once, \tool generates a small helper function with a local \lstinline|__int128 result| and places an XMM-based inline assembly block around \lstinline|result|. This combined construction triggers the release-build GCC-15 defect.

AddressSanitizer reports a heap-use-after-free in an array during register allocation. 
This array, \lstinline|update_hard_regno_preference|, stores register preferences and is accessed after being freed. The accessed range depends on the total number of registers tracked (hard and pseudo registers), 
with a theoretical upper bound of \lstinline|INT_MAX|, indicating that the UAF provides a large read/write space.

\textbf{Use-After-Poison:} LLVM-22 triggers a \texttt{use-after-poison} bug when compiling the C++ program in Listing~\ref{lst:conversion_operator_case}. This bug is triggered by two features:
\noindent
(1) a conversion operator returning a reference to a local object, e.g., \lstinline|operator const std::string&() const|, and
(2) a member whose type requires an intermediate conversion to produce the target reference type, e.g., \lstinline|std::shared_ptr<std::string>| to \lstinline|const std::string&|.

\vspace{6pt}
\begin{lstlisting}[language=C++, 
caption={LLVM-191687: AddressSanitizer use-after-poison during compilation.},
label=lst:conversion_operator_case,
basicstyle=\footnotesize\ttfamily,
columns=flexible
]
class A {
public:
    std::shared_ptr<std::string> s;
    // Feature1: Conversion operator returning 
    // reference to local object.
    operator const std::string&() const {
        // Feature2: Intermediate conversion.
        return *s; 
    }
};
void foo() {
    A a;
    const std::string& s = a;
}
\end{lstlisting}

To realize these two features together, \tool first defines a class 
\lstinline|A| that includes the 
\lstinline|operator const std::string&() const| conversion operator. 
Next, it introduces a local object of type \emph{shared pointer to std::string}, named \lstinline|s|, 
and the conversion operator returns \lstinline|*s|. 
This materializes the required conversion from \lstinline|std::shared_ptr<std::string>| to \lstinline|const std::string&|. 

AddressSanitizer reports a \texttt{use-after-poison} during compilation. The root cause is a heap out-of-bounds write caused by an incorrect internal offset when Clang constructs type information for a conversion-function declarator. As a result, \texttt{memcpy} writes 24 bytes beyond the intended heap object into poisoned memory. This yields a source-triggerable heap memory corruption bug during compilation.

\textbf{Assertion Failure:} GCC-15 crashes during interprocedural analysis (IPA) when compiling the program in Listing~\ref{lst:ipa_cases} with the \texttt{-O3} optimization flag. This bug is primarily triggered by two novel features introduced by \groupllm.
(1) a \texttt{static\_cast} to a base class type within a method, and
(2) calling a method through \texttt{static\_cast}.

\begin{lstlisting}[language=C++, 
caption={GCC-122856: An assertion failure occurs during interprocedural analysis in GCC-15, triggered by two novel features introduced by \groupllm.},
    basicstyle=\footnotesize\ttfamily,
commentstyle=\footnotesize\ttfamily\color{green!50!black},
    label=lst:ipa_cases,
    columns=flexible
]
struct B {
    virtual void foo();
};

// Feature1: A static_cast to a base class type within a method.
struct C : B {
    // Feature2: Call a method through static_cast.
    void foo() { static_cast<B*>(this)->foo(); }
};

int main() {
    C c;
    c.foo();
}
\end{lstlisting}

For the first feature, \tool constructs two structs, \texttt{struct B} and \texttt{struct C}, establishing an inheritance relationship where \texttt{struct C} inherits from \texttt{struct B}. 
Then, \tool converts an instance of \texttt{struct C} to its base class \texttt{struct B} by \texttt{static\_cast} at line 8 in Listing~\ref{lst:ipa_cases}.

For the second feature, in order to call a method through a \texttt{static\_cast}, \tool creates a virtual method \texttt{foo} for both \texttt{B} and \texttt{C}, and then calls \texttt{foo} using \texttt{static\_cast<B*>(this)->foo()}.

Consequently, this setup triggers an infinite recursion involving a virtual function and type casting.
The recursion process is as follows: \texttt{c.foo()} \(\rightarrow\) \texttt{static\_cast<B*>(this)->foo()} \(\rightarrow\) \texttt{c.foo()}.
Since \texttt{foo} is a virtual method, the actual call to \texttt{b.foo()} is resolved to \texttt{c.foo()}.
This recursion process involves the \texttt{static\_cast} and the \texttt{virtual method foo},  making it difficult for GCC to perform interprocedural analysis, ultimately triggering an internal assertion failure.
It is worth noting that these two features are introduced by \groupllm, which suggests that the new features introduced by \groupllm also have the potential to trigger bugs.

\noindent\textbf{Answer to RQ4.}
By combining features, \tool uncovers \numberReportedBug compiler defects with non-trivial security implications, including \numberReleaseBuildBug bugs in release compilers and \numberReleaseMemorySafetyBug memory-safety bugs.

\section{Related Work}

\textbf{Mutation-based Testing.}
Mutation-based fuzzers generate programs by applying a set of mutators to existing programs~\citeN{AnSurveyOfTheDevelopmentOfMutationTetsing2011, MutationSurvey2019}.
GrayC uses manually defined structural mutators~\cite{GrayC2023}, whereas
\metamut, \mutfall, and \issuemut use LLMs and bug reports to expand the mutation space~\citeN{MetaMut2024,Mut4All2025,Issue2mut2025}: MetaMut designs about 100 mutators, Mut4All derives 403 from bug reports, and IssueMut adds 587.
TyMut targets complex C++ types~\cite{TyMut2025}; EMI eliminates dead branches,
Athena removes dead branches or inserts code into unexecuted regions, and
Hermes injects dead snippets into live regions
~\citeN{EMIPLDI14,Athena2015,HermesOOPSLA2016}. These systems mutate syntax or
predefined structures to exercise specific optimizations. \tool instead
combines semantic conditions together with their dependencies.

\textbf{Function Synthesis Testing.}
Function synthesis composes real-world functions through shared state or calls.
\creal links open-source functions through global variables~\cite{Creal2024},
and \legofuzz expands the function pool with LLM agents~\cite{LegoFuzz2025}.
Both compose at function granularity and reuse realistic snippets with complex control and data flow. 
\tool instead represents relationships between arbitrary program elements as features, enabling finer-grained composition.

\textbf{Generation-based Testing.}
Generation-based techniques construct programs from scratch~\citeN{CompilerTestingSurvey2020, GrammarsForTestDataGeneration1990}.
Csmith uses a C grammar~\cite{CsmithPLDI11}; \yarpgen and YARPGen v.2 use
templates for complex control/data flow and loops~\citeN{YARPGen2020,YARPGenv22023};
All these techniques use manually written templates or rules to ensure the output programs exhibit the required semantics.
\tool captures semantics through features and uses LLMs to instantiate them, avoiding manually defined rules for program generation.

\textbf{LLM-based Generation.}
Leveraging the coding strengths of LLMs~\citeN{ge2025surveyvibecodinglarge, zhao2025systematicstudygeneratingweb}, many efforts employ LLMs for compiler testing~\cite{ASurveyOfModernCompilerFuzzing}.
\fuzzfall uses LLMs for generation and mutation~\cite{fuzz4all2024}; WhiteFox targets deep-learning compilers~\cite{WhiteFox2024}; and FuzzGPT and ClozeMaster fine-tune on code snippets~\cite{FuzzGPT2024,ClozeMaster2025}. 
Programs generated directly by these models can repeat structures from their training data~\cite{LegoFuzz2025,FuzzGPT2024}. 
\tool instead collects a large set of features, and the vastness of their combinatorial space enables the generation of more diverse programs.

\section{Discussion}

Extracted features may include relevant but non-essential conditions because their quality depends on the LLM, trigger-program conciseness, and the completeness of bug reports and fixes. 
Nevertheless, our evaluation shows that they remain effective for compiler testing.

\tool requires that the generated code satisfies all features specified in the feature groups, which demands coding ability from LLMs. 
\tool requires the model to understand and integrate diverse program semantics and dependencies. \tool enhances feature-group coherence through \groupllm, which simplifies the generation task. 
As model capabilities improve, or through fine-tuning, these challenges can be further alleviated, leading to more accurate code generation.

\section{Conclusion}
We present \tool, a feature-based fuzzing framework. \tool collects bug-prone features and combines them to generate programs for testing compilers. 
Our evaluation on GCC and LLVM shows that \tool improves both code coverage and bug-finding effectiveness, 
resulting in the discovery of \numberReportedBug real-world compiler bugs, \numberConfirmBug of which have already been confirmed by compiler developers.
\tool also demonstrates security relevance by uncovering \numberReleaseBuildBug release-build bugs, including \numberReleaseMemorySafetyBug memory-corruption bugs.

\begin{acks}
This work was supported in part by grants from the National Natural Science Foundation of China (62572231, 62172201, 62032010, 62432004, and 62302207),  the Fundamental Research Funds for the Central Universities 2026300278, ZZKT2026A31, the Fundamental and Interdisciplinary Disciplines Breakthrough Plan of the Ministry of Education of China (No. JYB2025XDXM118), and the “111 Center”(No. B26023).
\end{acks}

\bibliographystyle{ACM-Reference-Format}
\bibliography{sample-base}

\appendix

\section*{Ethical Considerations}

This work involves no user studies or personal, sensitive, or proprietary data;
all experiments use open-source compilers and generated programs. We
responsibly report confirmed findings through the official GCC and LLVM bug
trackers, follow established disclosure practices, and publish no weaponizable
exploits or proof-of-concept attacks. \tool uses LLMs only to extract,
compose, and instantiate compiler-test features, not for human decisions,
personal data, or social inference, and the generated programs target compiler
robustness rather than downstream systems or users. Although advanced fuzzing
can accelerate vulnerability discovery, our goal is to make compiler defects
easier to detect and repair, thereby reducing long-term risks to critical
software infrastructure.

\section*{Open Science and Artifact Availability}

Our artifacts are publicly available at
\url{https://github.com/addaddgo/FeatureFuzz}. They include \tool's Python
implementation for feature extraction, group
synthesis, program instantiation, fuzzing, coverage, and crash collection. All
evaluated data are derived from public bug reports, triggering programs, and
fixes from GCC Bugzilla and the official GCC and LLVM repositories, and we will
release the evaluated feature pool and groups, or a representative subset,
together with the ExtractLLM prompts; no private datasets are used. For the
public GCC and LLVM versions in our experiments, we will document the software
environment, compiler builds, and AFL++/gcov instrumentation, while hardware is
reported only for performance context. We will also release the system prompts
and scripts for constructing GroupLLM training data. The OpenAI-compatible
implementation supports OpenRouter, Aliyun, and local vLLM/HuggingFace
deployments, although exact reproduction may depend on checkpoint and provider
availability.


\begingroup
\small
\linespread{0.97}\selectfont
\setlength{\intextsep}{5pt plus 1pt minus 1pt}

\section{Supplementary Material}

\subsection{\groupllm Training Details}
\label{appendix:groupllm}

\groupllm is fine-tuned on feature groups extracted from historical bugs. We
group features by Bug ID so that every training example originates from a
program that exercised a coherent compiler behavior. Because the number of
features per bug is highly imbalanced, we construct the training pairs based on
group size. For two-feature groups, we use both feature orders. For
three-feature groups, we sample multiple input--target subsets. For groups with
at least four features, we generate four independently shuffled instances and
randomly split each instance into a non-empty input prefix and target suffix.
This procedure increases the representation of sparse groups while preventing
large groups from dominating the corpus.

Each pair is converted into a Role--Task--Input--Output example. The role asks
the model to act as a compiler-feature synthesis solver; the task asks it to
complete a partial feature set; the input contains the observed features; and
the output contains additional features that preserve their implicit control-
and data-flow relationships. The resulting corpus contains approximately
55,000 examples (91MB), of which 10\% are held out for validation.

\begin{table}[h]
  \centering
  \small
  \caption{Hyperparameters used to fine-tune \groupllm.}
  \label{tab:groupllm_training}
  \begin{tabular}{ll}
    \toprule
    \textbf{Configuration} & \textbf{Value} \\
    \midrule
    Base model & Qwen3-4B \\
    Adaptation & LoRA on all linear layers \\
    LoRA rank / $\alpha$ & 8 / 16 \\
    Optimizer & AdamW \\
    Peak learning rate & $5 \times 10^{-5}$ \\
    Schedule & Cosine decay \\
    Epochs / effective batch size & 3 / 16 \\
    Precision / maximum length & BF16 / 2048 tokens \\
    \bottomrule
  \end{tabular}
\end{table}

The loss initially drops from 2.35 to 1.40 and then converges to 1.04 over
11,829 optimization steps. During fuzzing, two sampled features are placed in
the input field, and \groupllm predicts the complementary features used to form
a coherent group.

\subsection{Semantic Coherence and Compilation Success}
\label{app:semantic-correlation}

\begin{figure}[h]
  \centering
  \includegraphics[width=0.9\columnwidth]{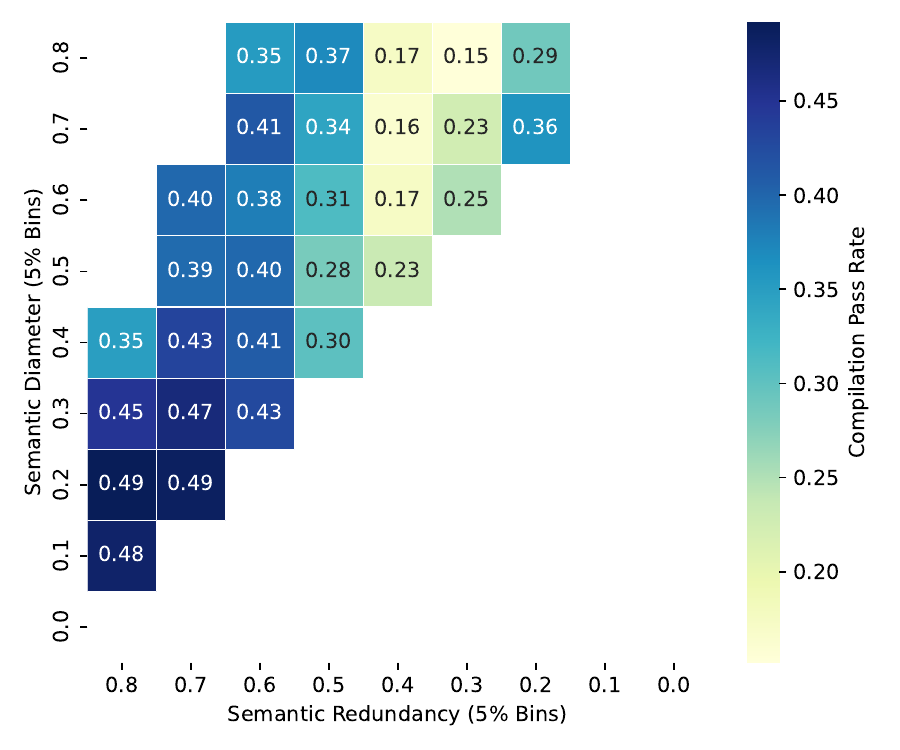}
  \vspace{-12pt}
  \caption{Compilation pass rate across semantic-redundancy and
  semantic-diameter bins.}
  \label{fig:semantic_compilation_pass_rate}
\end{figure}

Table~\ref{tab:semantic_stats_transposed} shows that groups synthesized by
\groupllm are closer to the coherent groups extracted from real programs than
groups produced by random sampling. Figure~\ref{fig:semantic_compilation_pass_rate}
provides the corresponding sample-level view. Compilation success is generally
higher when redundancy exceeds 0.6 and diameter remains below 0.4, indicating
that related features provide \instantiatellm with a shared program context.
Pass rates decline as diameter increases because distant features require more
complex integration. The small rebound in some high-diameter,
low-redundancy bins occurs when the model implements weakly related features
as largely independent code fragments; these programs can compile despite
providing weaker semantic integration.

\endgroup

\end{document}